\documentclass[prl,aps,reprint,superscriptaddress,floatfix,nofootinbib]{revtex4-2}

\usepackage[utf8]{inputenc}
\usepackage{graphicx}

\graphicspath{{./figures/}}

\usepackage{amsmath}
\usepackage{xspace}

\usepackage[dvipsnames]{xcolor}
\definecolor{dark-blue}{rgb}{0,0.2,0.6}

\usepackage[colorlinks,%
            pdfusetitle,%
            urlcolor=dark-blue,%
            citecolor=dark-blue,%
            linkcolor=dark-blue]{hyperref}

\usepackage{etoolbox}
\makeatletter
\pretocmd{\NAT@open}{\begingroup\color{\@citecolor}}{}{}
\apptocmd{\NAT@close}{\endgroup}{}{}
\makeatother

\newcommand{\yb}{\ensuremath{{^\text{173}\text{Yb}}}\xspace}
\newcommand{\sr}{\ensuremath{{^\text{87}\text{Sr}}}\xspace}

\newcommand{\sS}[1]{\ensuremath{{^1\text{S}_{#1}}}}

\newcommand{\asciimathunit}[1]{\ensuremath{\,\text{#1}}}

\newcommand{\nm}{\asciimathunit{nm}}
\newcommand{\mum}{\asciimathunit{\textmu m}}

\newcommand{\s}{\asciimathunit{s}}

\newcommand{\kB}{k_\text{B}}

\newcommand{\N}{\ensuremath{N}\xspace}

\newcommand\at[2]{\left.#1\right|_{#2}}

\newcommand{\subfigref}[2]{\hyperref[fig:#1]{\ref*{fig:#1}(#2)}}

\begin{document}


\title{Equation of State and Thermometry of the 2D SU(\N) Fermi-Hubbard Model}

\author{G.~Pasqualetti}\email{giulio.pasqualetti@lmu.de}
\author{O.~Bettermann}
\author{N.~\surname{Darkwah Oppong}}
\affiliation{Ludwig-Maximilians-Universit{\"a}t, Schellingstra{\ss}e 4, 80799 M{\"u}nchen, Germany}
\affiliation{Max-Planck-Institut f{\"u}r Quantenoptik, Hans-Kopfermann-Stra{\ss}e 1, 85748 Garching, Germany}
\affiliation{Munich Center for Quantum Science and Technology (MCQST), Schellingstra{\ss}e 4, 80799 M{\"u}nchen, Germany}

\author{E.~\surname{Ibarra-Garc{\'i}a-Padilla}}
\affiliation{Department of Physics and Astronomy, Rice University, Houston, Texas 77005-1892, USA}
\affiliation{Rice Center for Quantum Materials, Rice University, Houston, Texas 77005-1892, USA}
\affiliation{Department of Physics, University of California, Davis, California 95616, USA}
\affiliation{Department of Physics and Astronomy, San Jos\'e State University, San Jos\'e, California 95192, USA}

\author{S.~Dasgupta}
\affiliation{Department of Physics and Astronomy, Rice University, Houston, Texas 77005-1892, USA}
\affiliation{Rice Center for Quantum Materials, Rice University, Houston, Texas 77005-1892, USA}

\author{R. T. Scalettar}
\affiliation{Department of Physics, University of California, Davis, California 95616, USA}

\author{K. R. A. Hazzard}
\affiliation{Department of Physics and Astronomy, Rice University, Houston, Texas 77005-1892, USA}
\affiliation{Rice Center for Quantum Materials, Rice University, Houston, Texas 77005-1892, USA}
\affiliation{Department of Physics, University of California, Davis, California 95616, USA}

\author{I.~Bloch}
\author{S.~F{\"o}lling}
\affiliation{Ludwig-Maximilians-Universit{\"a}t, Schellingstra{\ss}e 4, 80799 M{\"u}nchen, Germany}
\affiliation{Max-Planck-Institut f{\"u}r Quantenoptik, Hans-Kopfermann-Stra{\ss}e 1, 85748 Garching, Germany}
\affiliation{Munich Center for Quantum Science and Technology (MCQST), Schellingstra{\ss}e 4, 80799 M{\"u}nchen, Germany}

\hypersetup{pdfauthor={G.~Pasqualetti, O.~Bettermann, N.~Darkwah Oppong, E.~Ibarra-Garc{\'i}a-Padilla, S.~Dasgupta, R. T. Scalettar, K. R. A. Hazzard, I.~Bloch, S.~F{\"o}lling}}

\date{May 30, 2023}

\begin{abstract}
We characterize the equation of state (EoS) of the SU($N>2$) Fermi-Hubbard Model (FHM) in a two-dimensional single-layer square optical lattice.
We probe the density and the site occupation probabilities as functions of interaction strength and temperature for $\N = 3, 4$ and 6.
Our measurements are used as a benchmark for state-of-the-art numerical methods including determinantal quantum Monte Carlo (DQMC) and numerical linked cluster expansion (NLCE).
By probing the density fluctuations, we compare temperatures determined in a model-independent way by fitting measurements to numerically calculated EoS results, making this a particularly interesting new step in the exploration and characterization of the SU(\N) FHM.
\end{abstract}

\maketitle


The interest in the square lattice SU(2) Fermi-Hubbard Model (FHM) has been historically driven by its suitability to describing cuprate superconductors, owing to their layered character and exceptionally simple band structure near the Fermi surface.
For other more complex and multi-orbital materials, however, descriptions with $\N > 2$ spin components have long been used, which, in addition to being of fundamental interest, provide an elegant approximation of degenerate orbitals using a higher symmetry group.
Larger \N systems, in particular in 2D geometries, are relevant for describing the physics of transition-metal oxides~\cite{tokuraOrbitalPhysicsTransitionMetal2000,dagottoColossalMagnetoresistantMaterials2001,liSU4TheorySpin1998}, orbitally-selective Mott transitions~\cite{demediciOrbitalselectiveMottTransition2005,florensQuantumImpuritySolvers2002,florensSlaverotorMeanfieldTheories2004,sprauDiscoveryOrbitalselectiveCooper2017}, graphene's SU(4) spin valley symmetry~\cite{goerbigElectronicPropertiesGraphene2011}, twisted-bilayer graphene~\cite{xuKekulValenceBond2018,youSuperconductivityValleyFluctuations2019,natoriSUHeisenbergModel2019a,daliaoCorrelationInducedInsulatingTopological2021}, the Kondo effect~\cite{nozieresKondoEffectReal1980,coxExoticKondoEffects1998}, heavy fermion behavior~\cite{hewsonKondoProblemHeavy1993}, and achieving robust itinerant ferromagnetism~\cite{katsuraNagaokaStates2013,bobrowExactResultsItinerant2018}.
The SU(\N) FHM is a special case of the $N>2$ models that enjoys a higher symmetry group that stabilizes quantum fluctuations~\cite{affleckLargenLimitHeisenbergHubbard1988}, making it a fertile ground for theory, and constituting a baseline to more complex multi-orbital models.
The determination of the $\N>2$ equation of state (EoS) of the SU(\N) FHM is an important milestone in the attempt of understanding its properties.
However, the exponential scaling of the Hilbert space with \N and the increased severity of the fermion sign problem~\cite{batrouniAnomalousDecouplingsFermion1990} make its numerical simulation more challenging than the $\N=2$ case~\cite{assaadPhaseDiagramHalffilled2005,ibarra-garcia-padillaUniversalThermodynamicsMathrmSU2021a,leeFillingdrivenMottTransition2018,ibarra-garcia-padillaThermodynamicsMagnetismTwodimensional2020}.

Ultracold atoms in an optical lattice have provided valuable quantum simulations of the SU(2) FHM~\cite{grossQuantumSimulationsUltracold2017}.
They complement and can sometimes outperform classical simulations~\cite{bohrdtExplorationDopedQuantum2021a,daleyPracticalQuantumAdvantage2022}.
More recently, the SU($\N>2$) FHM has been successfully explored with ultracold alkaline-earth-like atoms such as \yb or \sr in optical lattices, which naturally feature a full SU(\N) symmetry in the atomic ground state~\cite{wuExactSymmetrySpin2003,honerkampUltracoldFermionsSU2004,cherngSuperfluidityMagnetismMulticomponent2007,cazalillaUltracoldGasesYtterbium2009,hermeleMottInsulatorsUltracold2009,gorshkovTwoorbitalSUMagnetism2010,sotnikov2014,sotnikov2015,unukovych2021}.
A substantial effort has been placed in probing the thermodynamics and the short-range correlations of the model for different spin degeneracies and lattice geometries, and experiments have gone well-beyond the regime that can be calculated with theory~\cite{taieSUMottInsulator2012b,hofrichterDirectProbingMott2016,ozawaAntiferromagneticSpinCorrelation2018,tusiFlavourselectiveLocalizationInteracting2022,abelnInterorbitalInteractionsSU2021,taieObservationAntiferromagneticCorrelations2022,singh2022}.
However, the SU(\N) generalization remains much less explored and understood compared to the SU(2) case~\cite{takahashiQuantumSimulationQuantum2022}.
This is particularly true in two dimensions, where the thermodynamics of the SU(2) FHM at intermediate temperatures have been studied extensively~\cite{cocchiEquationStateTwoDimensional2016,drewesThermodynamicsLocalDensity2016,drewesAntiferromagneticCorrelationsTwoDimensional2017,cocchiMeasuringEntropyShortRange2017a,bollSpinDensityresolvedMicroscopy2016,hilkerRevealingHiddenAntiferromagnetic2017,koepsellImagingMagneticPolarons2019,koepsellMicroscopicEvolutionDoped2021,greifSiteresolvedImagingFermionic2016,chiuStringPatternsDoped2019,jiCouplingMobileHole2021,mazurenkoColdatomFermiHubbard2017a,parsonsSiteresolvedMeasurementSpincorrelation2016,cheukObservation2DFermionic2016,cheukObservationSpatialCharge2016,nicholsSpinTransportMott2019,hartkeDoublonHoleCorrelationsFluctuation2020}.

In this Letter, we probe the equation of state of the two-dimensional SU(\N) FHM in a square lattice at intermediate temperatures in both the metallic and the Mott regime and compare our results with numerical calculations. 
In particular, we determine the in-lattice temperature and entropy by fitting experimental data using numerical methods such as determinantal quantum Monte Carlo (DQMC)~\cite{blankenbeclerMonteCarloCalculations1981,sorellaNovelTechniqueSimulation1989} and numerical linked cluster expansion (NLCE)~\cite{rigolNumericalLinkedCluster2006,tangShortIntroductionNumerical2013}. 
We additionally determine the entropies in the 2D bulk before loading and after unloading from the lattice potential, and separately characterize the system inside the lattice with a thermometry relying on the fluctuation-dissipation theorem (FDT) based on the measurement of density fluctuations, without requiring modeling by theory.

\begin{figure}[!t]
	\includegraphics[width=\columnwidth]{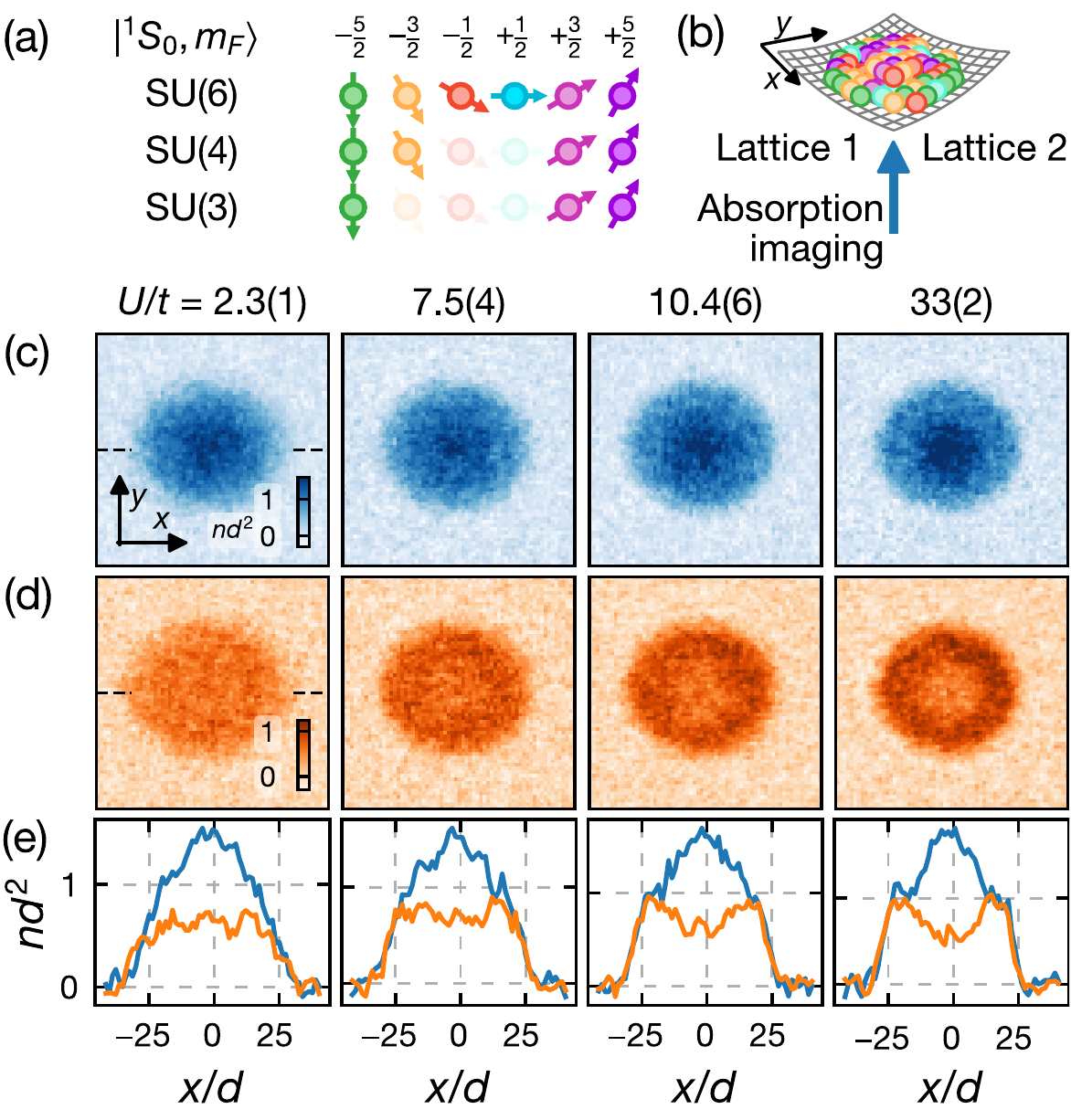}
	\caption{\label{fig:spatial}%
  Probing the 2D SU(\N) Fermi-Hubbard model with ultracold atoms.
  (a)~The $\sS{0}$ ground state of \yb naturally features an SU(6) symmetry, which can be freely tuned to $N\leq 6$ by preparing a suitable combination of the nuclear spin states $m_F = -5/2, -3/2, \dots, +5/2$ (colored circles and arrows).
  (b)~Schematic of the experimental setup showing a gas in a single layer 2D square lattice with harmonic confinement detected with absorption imaging along gravity.
	(c)~Spatial distribution of the density $n(x,y)$ for $\N = 6$.
  Each cloud shown in the horizontal frame has been prepared with the same initial entropy in the bulk and loaded into the lattice to a different $U/t$ value. 
	(d)~Singly-occupied sites after parity projection.
  Each horizontal frame corresponds to the same state shown in the same column of (c).
	(e)~Density profiles for the data shown in (c) and (d) along the corresponding dashed lines in the left frames.
  Each image was produced using the averaging of 8 shots after center of mass alignment~\cite{SM}.
  }
\end{figure}
%

The SU(\N) FHM Hamiltonian is given by:
\begin{equation}
  \hat H = -t \sum_{\langle i,j\rangle,\sigma}{\left(\hat c_{i\sigma}^\dagger \hat c_{j\sigma}^{\phantom\dagger} + \text{h.c.}\right)} + \frac{U}{2}\sum_{i,\sigma\neq\tau}{\hat n_{i\sigma}\hat n_{i\tau}} - \sum_{i,\sigma}{\mu_i\hat n_{i\sigma}},
  \label{eq:SUN_FHM}
\end{equation}
where $\hat c_{i\sigma}^\dagger$ and $\hat c_{i\sigma}^{\phantom\dagger}$ represent the fermionic creation and annihilation operators at site $i$ with spin $\sigma \in \{1\dots \N\}$,
$\hat n_{i\sigma} = \hat c_{i\sigma}^\dagger \hat c_{i\sigma}^{\phantom \dagger}$ is the number operator,
$\langle i,j\rangle$ denotes next-neighbor lattice sites, $t$ is the hopping amplitude, $U$ is the on-site interaction strength and $\mu$ denotes the chemical potential, which absorbs the contribution of the trap confinement in the local density approximation (LDA)~\cite{nascimbeneExploringThermodynamicsUniversal2010}.

In this work, we directly probe the local density, components of the site-occupation distribution, and the density fluctuations within the detection resolution of a few lattice sites.
By differentiating the density with respect to the local chemical potential, we evaluate the isothermal compressibility $\kappa = \at{\partial n/\partial \mu}{T}$.
Crucially, we implement a 2D single-layer SU(\N) ensemble that we probe with perpendicular high-resolution absorption imaging.
This avoids integrating over inhomogeneous stacks of 2D systems~\cite{darkwahoppongObservationCoherentMultiorbital2019a,darkwahoppongProbingTransportSlow2022} and allows us to directly access the density profile without complex reconstruction techniques required in 3D~\cite{hofrichterDirectProbingMott2016} and access density fluctuations as an additional thermodynamic in-situ observable.

In our experiment, we start by loading a degenerate Fermi gas of \yb with tunable $\N \leq 6$ equally populated components [see Fig.~\subfigref{spatial}{a}] and an entropy per particle $s/\kB \gtrsim 1.0$ into the single, horizontal layer of a vertical lattice. 
In this layer, we adiabatically ramp up a 2D square lattice potential with a wavelength of $\lambda = 759\nm$ and a spacing of $d=\lambda/2$ [see Fig.~\subfigref{spatial}{b}]. 
By modifying the lattice depth, we can tune the strength of the interactions.
We measure the density distribution using in-situ, saturated absorption imaging with a spatial resolution of approximately $2\mum \approx 5d$~\cite{SM}.

The measured 2D density $n(x,y)$ of an SU(6) ensemble is shown in Fig.~\subfigref{spatial}{c} for different interaction strengths and the same initial state preparation in the 2D bulk (the potential without in-plane lattices).
Because of the harmonic confinement generated by the Gaussian profile of the lattice beams, the chemical potential varies across the trap, sampling different regions of the EoS.
For increasing interactions, and in particular when the on-site interaction is larger than the square lattice bandwidth ($U/t \gtrsim 8$), we observe the emergence of plateaus at integer density which we associate with an incompressible regime, a signature of a Mott insulating state.

\begin{figure*}[t]
	\includegraphics[width=\linewidth]{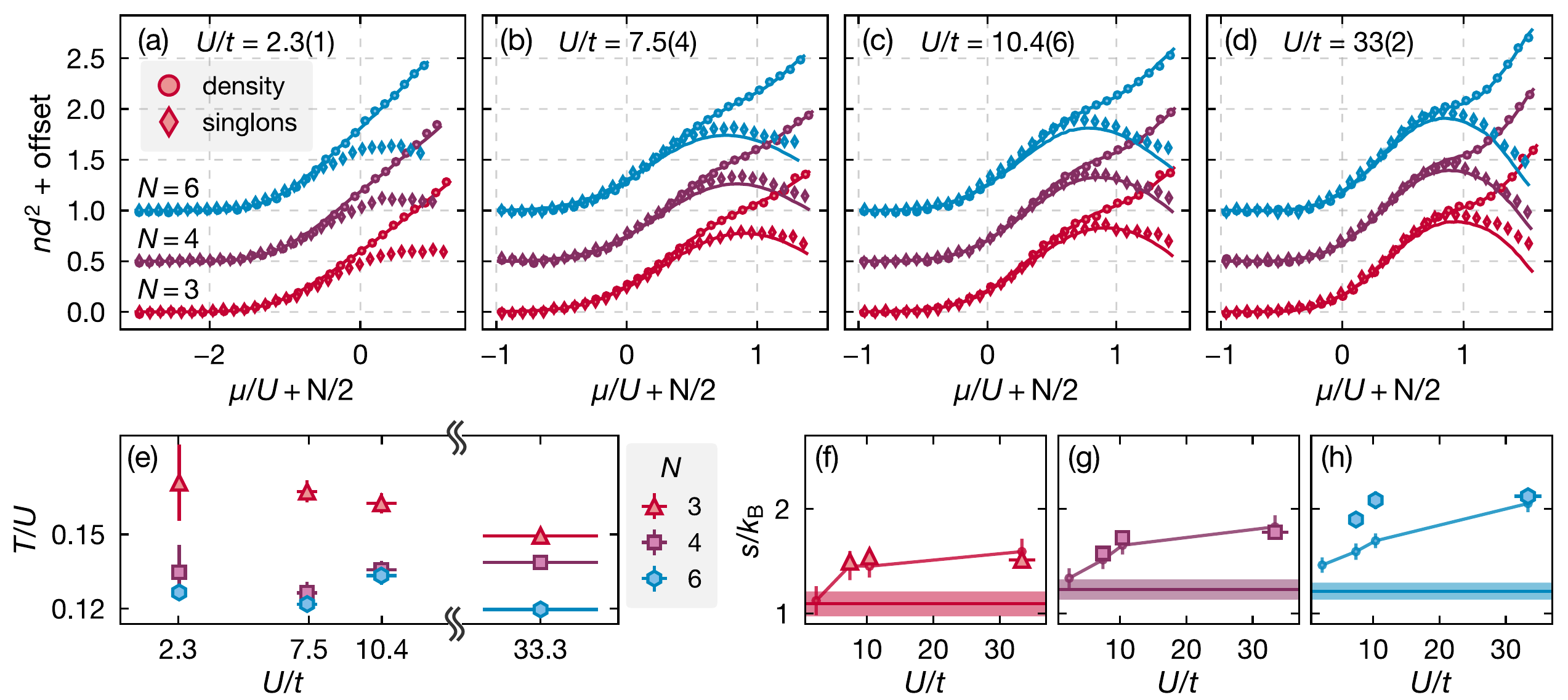}
	\caption{\label{fig:EoS}%
  Equation of state for the SU(\N) Fermi-Hubbard Model with $\N=6$ (blue), $\N=4$ (purple) and $\N=3$ (red). 
  (a)-(d)~Density (circles) and singly-occupied sites (diamonds) as a function of the chemical potential. 
  Data for $\N=4$ and $\N=6$ have been offset by $0.5\,nd^2$ and $1.0\,nd^2$ along the vertical axis, respectively. 
  Continuous lines associated to the density curves correspond to the fit of the EoS calculations to the total density as described in the text. 
  The theory used for the fit is DQMC for $U/t = 2.3(1)$ and NLCE for the other values of $U/t$.
  The results from this fit model are also used to calculate the expected pair / single site distribution measurement.
  The chemical potential is defined with respect to the reference half filling [$nd^2(\mu=0) = \N/2$].
  For each $U/t$, we fit the average of 15 frames with similar atom number after center of mass alignment~\cite{SM}.
  Error bars are the standard error of the mean (s.e.m.).
  (e)~Temperature according to the fit of the EoS shown in Fig.~\subfigref{EoS}(a-d).
  (f)-(h)~Entropy per particle. 
  Horizontal line: entropy in the 2D bulk before loading into the lattice; triangles, squares, hexagons: entropy in the lattice according to the fit of the EoS; small circles: entropy in the 2D bulk after a round-trip experiment.
  The entropy in the bulk takes into account the effect of the interactions and the 3D anisotropic density of states~\cite{SM}. 
  Error bars correspond to the s.e.m. of the fit results.
  }
\end{figure*}
%

As a distinctive probe of number squeezing effects in and close to the Mott regime, we determine the occupation number distribution by measuring the parity-projected density.
After tuning $U/t$, we freeze the motion of the atoms by rapidly increasing the lattice depth and applying a photoassociation beam~\cite{taieSUMottInsulator2012b}, which converts on-site pairs into excited-state molecules that are subsequently lost.
The process removes $>99\%$ of the on-site pairs and $\approx 5\%$ of the remaining atoms~\cite{SM}.
Fig.~\subfigref{spatial}{d} shows the distribution of the singly-occupied sites corresponding to the same states of Fig.~\subfigref{spatial}{c}.
The increase in depletion in the center with increasing interaction strength is a consequence of number squeezing to a high atom pair fraction.

To access different spin degeneracies, we prepare $\N < 6$ ensembles by removing spin components using optical pumping~\cite{SM}.
In Fig.~\subfigref{EoS}{a-d}, we show the EoS as a function of the local chemical potential $\mu/U$ for $\N \in \{6,4,3\}$.
The chemical potential at a given location is calculated from the potential of the trap $\mu(x,y)=\mu_0 - \frac{1}{2}\left(\kappa_x x^2 +\kappa_y y^2\right)$.
The exact shape of the potential is determined by fitting the density, where the trap frequencies are left as free parameters.
We use a combined fit of the densities for $\N = 3, 4$ and 6 for each separate $U/t$, but verify that separate fits for each $\N$ return values compatible with those of the combined fit.
The fit of the EoS is performed in two dimensions, leaving as free parameters the temperatures $T(U/t, \N)$ and the chemical potential $\mu_0(U/t, \N)$ at the center of the trap.
The theoretical density is convolved with the reconstructed point spread function (PSF)~\cite{SM} to take into account the imaging imperfections.

For the EoS of Fig.~\ref{fig:EoS}, each spin mixture has been prepared with the same initial entropy per particle $s/\kB = 1.2(1)$ in the 2D bulk before ramping up the lattice. 
We fit and benchmark NLCE and DQMC~\cite{SM} which are commonly used state-of-the-art methods for finite-temperature SU(2) Hubbard models in the regime we are considering but have only recently been extended and applied to the SU(\N) experimental regime, which requires calculations away from $n d^2 = 1$~\cite{ibarra-garcia-padillaThermodynamicsMagnetismTwodimensional2020,taieObservationAntiferromagneticCorrelations2022}.
This is, to our knowledge, the first application of SU(\N) NLCE to non-integer filling, and to the calculation of the occupancy distributions. Moreover, compared to previous works, the calculation has been extended to higher orders~\cite{SM} to ensure a better convergence at low temperatures.
For $U/t = 7.5(4)$ and $10.4(6)$ we fit both DQMC and NLCE, observing an excellent agreement between the theory~\cite{SM} and the experiment.
For $U/t = 33(2)$, we use NLCE and a high-temperature series expansion (HTSE-2), observing also in this case an excellent agreement~\cite{SM}.
For $U/t = 2.3(1)$, the temperature lies below the range of convergence of NLCE and we resort to DQMC alone.
In Fig.~\subfigref{EoS}{a-d}, for the cases in which we fit more than one model, we only plot the NLCE results, because the lines would overlap.

In addition to the total density, in Fig.~\subfigref{EoS}{a-d} we also characterize the distribution of on-site occupation numbers by removing doublons using the pair removal process described above.
Experimental measurements (diamonds) are compared with the NLCE prediction (lines) based on the fit of the density, without additional free fit parameters, and agree well with the experimental data whenever available. 
As opposed to the $\N=2$ case, where only double occupancies are allowed, higher occupancies occur for $\N>2$. 
Although the numbers of these occupancies are small for the results considered at the temperatures and chemical potentials presented here, the photoassociation technique can be used to probe triple occupancies and their dynamics~\cite{wernerSpectroscopicEvidenceEngineered2022}.

The harmonic confinement of the trap returned by the density fit can be compared to the confinement obtained from an independent measurement of the oscillatory motion of the atoms in the combined dipole potentials~\cite{SM}.
We find a discrepancy between about $13\,\%$ for $U/t = 7.5(4)$ and $40\,\%$ for $U/t = 33(2)$, which is not fully explained by tolerances or the trap loading model.
A possible contribution could be a lack of adiabaticity during the loading into the lattice~\cite{bonnesAdiabaticLoadingOneDimensional2012,natuLocalGlobalEquilibration2011,dolfiMinimizingNonadiabaticitiesOpticallattice2015}.
However, neither varying the speed of the lattice ramps up to a factor of four (up to $1\,\s$ length) nor variations of the atom number lead to significant changes in the fit results. 
This would require the non-adiabatic effects to produce minimal changes in density and parity profiles~\cite{SM}. 

In Fig.~\subfigref{EoS}{e} we plot the temperatures obtained by the fits of the EoS.
We observe a smaller temperature for larger \N, a behavior expected due to the Pomeranchuk effect~\cite{cazalillaUltracoldGasesYtterbium2009,hazzardHightemperaturePropertiesFermionic2012}, but somewhat weaker than the ideal theoretical predictions~\cite{hazzardHightemperaturePropertiesFermionic2012,ibarra-garcia-padillaUniversalThermodynamicsMathrmSU2021a} with the temperatures for $\N = 4$ and 6 differing from each other by up to $20\,\%$.
We interpret this as a consequence of the heating not depending on \N during the loading process, resulting in different entropies in the lattice.
This is supported by the results presented in Fig.~\subfigref{EoS}{f-h}.
We find that, despite the initial entropy in the 2D bulk before loading into the lattice being independent of \N, the entropy returned by the fit of the EoS is larger for larger \N, which explains the weakening of the Pomeranchuk effect.
We also determine the entropy per particle in the 2D bulk after a round-trip experiment, which adds an inverted ramp back to the 2D bulk system.
In this case, we obtain entropies comparable to those reported by the fit in the lattice for $\N = 3$ and $\N=4$ but smaller for $\N = 6$, similar to previous observations~\cite{hofrichterDirectProbingMott2016} and potentially indicating nonadiabatic effects in the preparation or return ramp.

\begin{figure}[t]
  \includegraphics[width=\columnwidth]{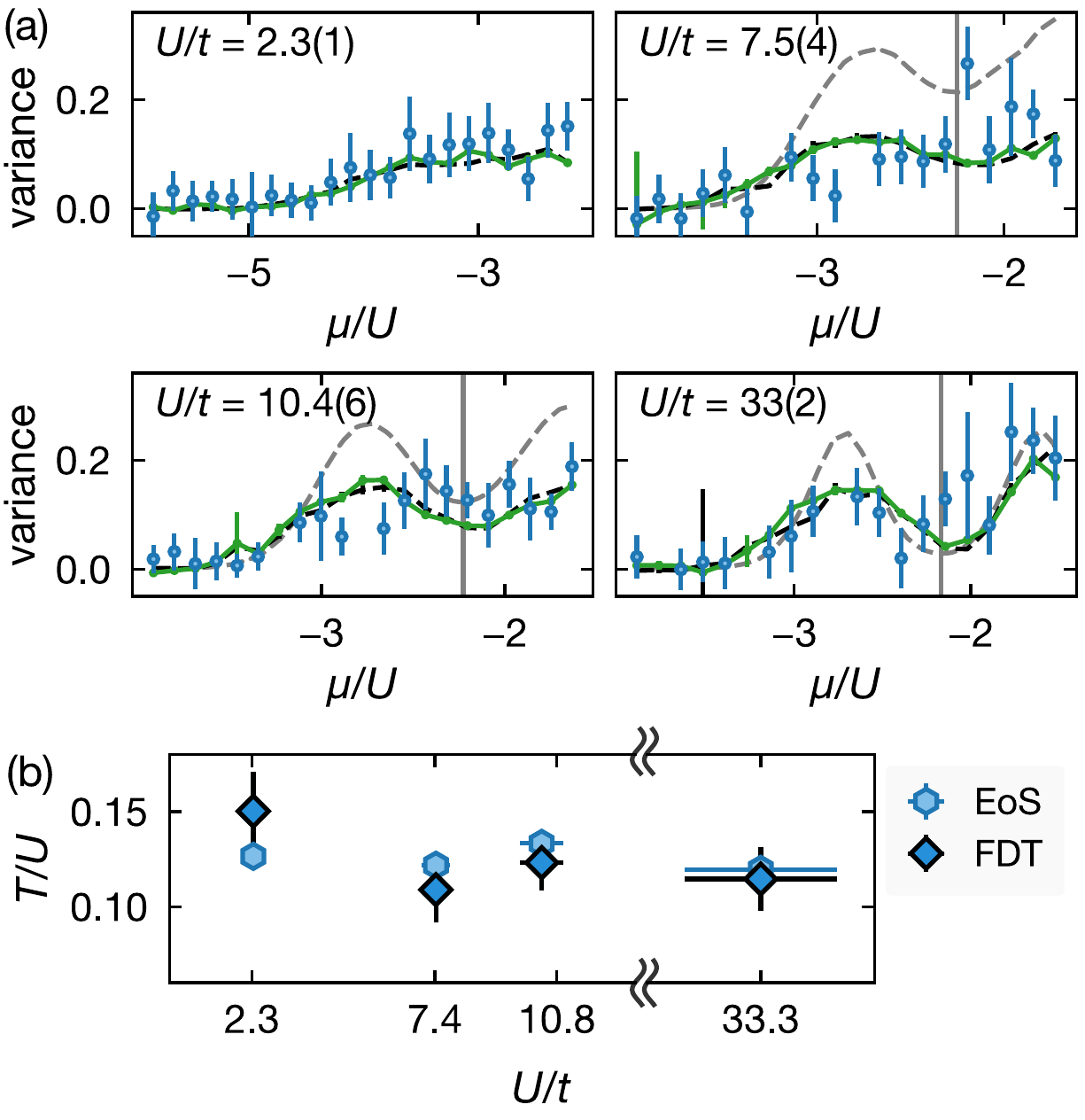}
  \caption{\label{fig:fluctuations} 
  (a)~Measured density fluctuations (blue) for $\N=6$ as a function of the chemical potential for different interaction strengths.
  The data points have been obtained from the variance of 15 frames (same as Fig.~\ref{fig:EoS}) computed on spatially-binned probe areas of size $\approx 5.1 \times 5.1\, d^2$ ($4 \times 4$ square camera pixels).
  The photon shot noise has been subtracted and a PSF correction has been taken into account~\cite{SM}.
  The green line corresponds to the numerically-differentiated compressibility $\kappa$ times the temperature $T_\text{EoS}$ obtained from the EoS-fit of the averaged data, while the black dashed line corresponds to the theory-derived compressibility times $T_\text{EoS}$.
  The vertical line indicates $\mu (nd^2 =1)$.
  The grey dashed line corresponds to the on-site density fluctuations $\delta n_0^2 = \langle \hat n^2\rangle - \langle \hat n\rangle^2$ calculated with NLCE for $T_\text{EoS}$. 
  (b)~Comparison of the temperatures $T_\text{FDT}$ (dark blue diamonds) and $T_\text{EoS}$ (light blue hexagons).
  Error bars are the s.e.m.
  }
\end{figure}
%

Complementary to the measurement of the EoS, the new possibility to directly access the density in the 2D SU(\N)-system allows us to probe the density fluctuations.
For an integration area of size $A\gg d^2$, the variance of the detected atom number is related to the isothermal compressibility $\kappa$ and the temperature $T$ through the fluctuation-dissipation theorem (FDT)~\cite{zhouUniversalThermometryQuantum2011}:
\begin{equation}
\text{var}\left(\int_A n\,dA\right) = \kB T \kappa A = \kB T A \at{\frac{\partial n}{\partial \mu}}{T}.
  \label{eq:FDT}
\end{equation}
By measuring the density fluctuations, we can access the temperature with spatial resolution, largely independently from the EoS models and without relying on fits of the potential parameters and to the theory~\cite{hartkeDoublonHoleCorrelationsFluctuation2020}.
In Fig.~\subfigref{fluctuations}{a} we show such density fluctuations as a function of the chemical potential for different $U/t$ values and $\N=6$.
For strong interactions, we observe a reduction of the fluctuations in the proximity of $nd^2=1$, where we expect an incompressible Mott-insulating regime.
Notably, the fluctuation amplitude is determined by area integration as described in Eq.~\eqref{eq:FDT}, and therefore agrees with the thermodynamic fluctuations from the FDT as opposed to the expected on-site fluctuations $\delta \hat n_0^2 = \langle \hat n^2 \rangle - \langle \hat n \rangle^2$ (grey dashed line). 
This discrepancy illustrates the role of non-vanishing short-range density correlations.

The FDT holds locally for each density.
In thermal equilibrium, the ratio between the fluctuations and the compressibility is constant.
We use the FDT to check this assumption and extract the temperature of the system.
For this purpose, we determine the isothermal compressibility $\kappa(\mu)$ directly from the density profile data with three-point differentiation and fit the temperature $T_\text{FDT}$ as the proportionality factor between the fluctuations and the compressibility.
In Fig.~\subfigref{fluctuations}{b} we compare $T_\text{FDT}$ (diamonds) with the temperature $T_\text{EoS}$ (hexagons) returned by the fit of the EoS.
We observe a good agreement for all interactions.
Moreover, we see that the residuals of the FDT analysis typically show similar temperatures at the center and at the edge of the cloud, indicating that there are no strong deviations from thermal equilibrium~\cite{SM}.

In conclusion, we report the measurement of the equation of state of the 2D SU(\N) FHM across the Mott crossover for temperatures comparable with or below the hopping energy and we compare the experiment with state-of-the-art numerical models. 
Moreover, with direct access to a single 2D plane system, we can independently determine temperatures in the experiment with spatial resolution using density fluctuation analysis, which allows one to e.g. cross-check thermal equilibrium.
This measurement characterizes the EoS also in regimes hard to reach by current numerical methods.
When compared to the experimental data, we find the theoretical calculations describe well the properties of the SU(\N) gas for the applicable range of temperatures.
The temperature measurements also indicate that thermal equilibration is not inhibited even in the case of deep lattices in a temperature range where the onset of spin correlations between sites is expected. 

The implementation of the directly-accessible 2D ensemble, together with the accompanying theoretical description, paves the way towards more direct quantum simulation of the typically-2D models of interest in naturally occurring systems with SU($\N>2$) representations such as transition metal oxides and orbitally-selective Mott transitions. 
An intriguing example is the case of cerium volume collapse, where there is a long-standing debate whether the single orbital Hubbard model ($\N=2$) or the double-orbital Hubbard model ($\N=4$)~\cite{johanssonAgTransitionCerium1974,allenKondoVolumeCollapse1982,lippAnomalousElasticProperties2017,heldCeriumVolumeCollapse2001} is the correct description. 
While in the condensed matter examples the SU(\N) symmetry is typically only approximately realized, cold atom representations provide an essentially exact realization of SU(\N), allowing to implement fully SU(\N)-symmetric and previously purely theoretical models.
It should even be possible to smoothly connect both regimes in a continuous way by controlled symmetry breaking using e.g. optical state manipulation or state-dependent potentials~\cite{tusiFlavourselectiveLocalizationInteracting2022,yi2008,gorshkovTwoorbitalSUMagnetism2010}, but more generally alkaline-earth-like quantum simulations of SU(\N) FHM can provide insight into the validity of the SU(\N) approximation in more realistic models.


\begin{acknowledgments}
We thank Alexander Impertro for his contributions in the early phase of the experiment. 
We thank Hao-Tian Wei for useful conversation and the exact diagonalization code used in the NLCE.
N.D.O. acknowledges funding from the International Max Planck Research School for Quantum Science and Technology.
E.I.G.P. is supported by the grant DE-SC-0022311, funded by the U.S. Department of Energy, Office of Science, and acknowledges support from  the Robert A. Welch Foundation (C-1872), the National Science Foundation (PHY-1848304). 
K.H. acknowledges support from the Robert A. Welch Foundation (C-1872) and the National Science Foundation (PHY-1848304), and the W. F. Keck Foundation (Grant No. 995764). 
Computing resources were supported in part by the Big-Data Private-Cloud Research Cyberinfrastructure MRI-award funded by NSF under grant CNS-1338099 and by Rice University's Center for Research Computing (CRC). 
K.H.'s contribution benefited from discussions at the Aspen Center for Physics, supported by the National Science Foundation grant PHY1066293, and the KITP, which was supported in part by the National Science Foundation under Grant No. NSF PHY1748958.
R.T.S. is supported by the grant DOE DE-SC0014671 funded by the U.S. Department of Energy, Office of Science. 
\end{acknowledgments}

\bibliography{references,sm}


\end{document}



\title{{\small Supplemental Material} \\Equation of State and Thermometry of the 2D SU(\N) Fermi-Hubbard Model}

\author{G.~Pasqualetti}\email{giulio.pasqualetti@lmu.de}
\author{O.~Bettermann}
\author{N.~\surname{Darkwah Oppong}}
\affiliation{Ludwig-Maximilians-Universit{\"a}t, Schellingstra{\ss}e 4, 80799 M{\"u}nchen, Germany}
\affiliation{Max-Planck-Institut f{\"u}r Quantenoptik, Hans-Kopfermann-Stra{\ss}e 1, 85748 Garching, Germany}
\affiliation{Munich Center for Quantum Science and Technology (MCQST), Schellingstra{\ss}e 4, 80799 M{\"u}nchen, Germany}

\author{E.~\surname{Ibarra-Garc{\'i}a-Padilla}}
\affiliation{Department of Physics and Astronomy, Rice University, Houston, Texas 77005-1892, USA}
\affiliation{Rice Center for Quantum Materials, Rice University, Houston, Texas 77005-1892, USA}
\affiliation{Department of Physics, University of California, Davis, California 95616, USA}
\affiliation{Department of Physics and Astronomy, San Jos\'e State University, San Jos\'e, California 95192, USA}

\author{S.~Dasgupta}
\affiliation{Department of Physics and Astronomy, Rice University, Houston, Texas 77005-1892, USA}
\affiliation{Rice Center for Quantum Materials, Rice University, Houston, Texas 77005-1892, USA}

\author{R. T. Scalettar}
\affiliation{Department of Physics, University of California, Davis, California 95616, USA}

\author{K. R. A. Hazzard}
\affiliation{Department of Physics and Astronomy, Rice University, Houston, Texas 77005-1892, USA}
\affiliation{Rice Center for Quantum Materials, Rice University, Houston, Texas 77005-1892, USA}
\affiliation{Department of Physics, University of California, Davis, California 95616, USA}

\author{I.~Bloch}
\author{S.~F{\"o}lling}
\affiliation{Ludwig-Maximilians-Universit{\"a}t, Schellingstra{\ss}e 4, 80799 M{\"u}nchen, Germany}
\affiliation{Max-Planck-Institut f{\"u}r Quantenoptik, Hans-Kopfermann-Stra{\ss}e 1, 85748 Garching, Germany}
\affiliation{Munich Center for Quantum Science and Technology (MCQST), Schellingstra{\ss}e 4, 80799 M{\"u}nchen, Germany}

\hypersetup{pdfauthor={G.~Pasqualetti, O.~Bettermann, N.~Darkwah Oppong, E.~Ibarra-Garc{\'i}a-Padilla, S.~Dasgupta, R. T. Scalettar, K. R. A. Hazzard, I.~Bloch, S.~F{\"o}lling}}

\date{May 30, 2023}

\maketitle

\renewcommand{\thefigure}{S\arabic{figure}}
\renewcommand{\theequation}{S.\arabic{equation}}
\renewcommand{\thesection}{S.\Roman{section}}
\renewcommand{\thetable}{S\arabic{table}}

{\small \tableofcontents}

\section{Experimental techniques}

\subsection{State preparation}
Our experiment starts by loading a spin-balanced unpolarized mixture ($\N=6$) of approximately $1.6 \times 10^6$ \yb atoms from a magneto-optical trap into a crossed optical dipole trap (XODT), where we perform forced evaporation to $T/T_\text{F}^\text{3D} < 0.2$.
We then perform a second stage of evaporation with an optical gradient, leading to an ensemble of $N_p \approx \num{2e3}$ atoms in the central plane of a vertical lattice with wavelength $\lambda = \SI{759.3}{nm}$ and lattice spacing $d_\text{vert} = \SI{3.9(3)}{\micro m}$.
The vertical band gap is \SI{3.95(1)}{kHz}, determined from a measurement of the $0\rightarrow2$ band resonance with parametric modulation.
In this configuration, the coefficients of the harmonic confinement are $(\kappa_x, \kappa_y)d^2 = [\num{1.35(3)}, \num{2,23(6)}]\times h$.
Numerical simulations~\cite{impertroPreparationStudy1D2020} predict that for $N_p \lesssim \num{5e3}$, nearly all the atoms should be loaded in the single central plane of the lattice.
We verify this assumption with a momentum refocussing technique similar to the one described in Ref.~\cite{hueckTwoDimensionalHomogeneousFermi2018a}. 

Mixtures with $\N=4$ are prepared in the XODT before the evaporation by optically pumping the population of the two nuclear spin states $m_F \in \{\pm 1/2 \}$ to the other spin states with four circularly-polarized pulses on the $\sS{0}\rightarrow\tP{1}$ intercombination line at \SI{50}{G}.
Mixtures with $\N=3$ are prepared in a similar fashion by pumping the spin components $m_F \in \{\pm 1/2, -3/2\}$ to the other states with five pulses.
In both cases, we check the balancing of the spin components with an Optical Stern-Gerlach (OSG) technique in time of flight~\cite{stellmer2011}. 
The standard deviation of the population of the selected spin components is below $\SI{5}{\percent}$ per component for SU(6) and SU(3) and below $\SI{8}{\percent}$ for SU(4).
The residual fraction of unwanted spin components is below \SI{5}{\percent} per component.

\subsection{Lattice loading and Hubbard parameters}
\label{sec:lattice_loading}

\begin{figure}[t]
  \includegraphics[width=\columnwidth]{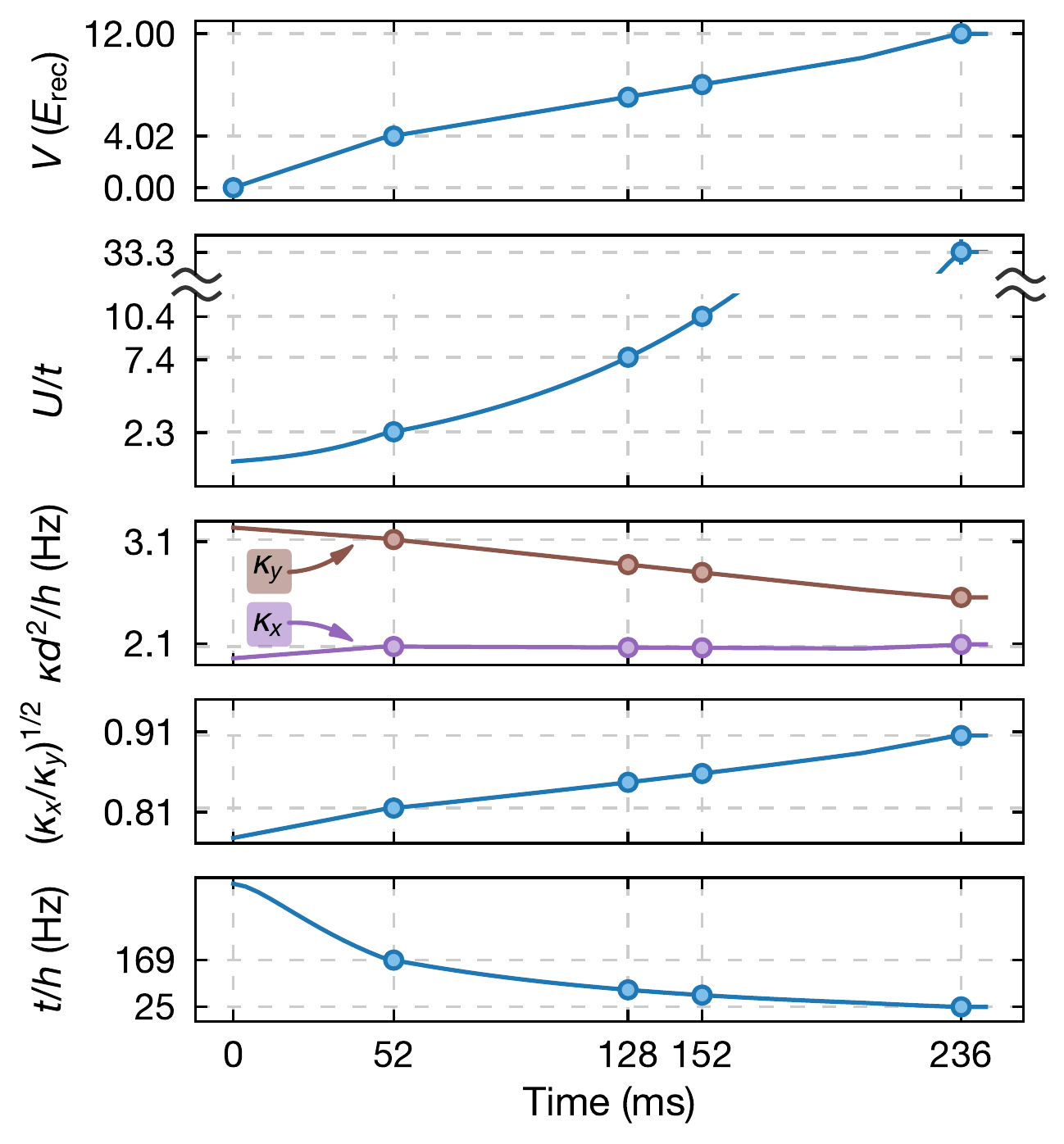}
  \caption{\label{fig:sequence_params} 
  Time dependence of the lattice depth $V$ and the Hubbard parameters $U$, $\kappa_x$, $\kappa_y$ and $t$ during the lattice ramps.
  The lattice depth and the hopping along the spatial $x$ and $y$ directions are the same.
  }
\end{figure}

The two orthogonal in-plane lattices have the same depth and they are tuned to the desired value following the ramp profile shown in Fig.~\ref{fig:sequence_params}.
The ramp speed has been chosen to minimize the entropy in the round-trip experiment.
After reaching the desired lattice depth, we quickly switch off the vertical lattice to avoid light-assisted collisions and to perform in situ imaging.

\begin{table}[t]
\begin{ruledtabular}
\begin{tabular}{SSSSSS}
{\footnotesize $U/t$} & {\footnotesize \text{$V$ ($\Erec$)}} & {\footnotesize \text{$U/h$ (Hz)}} & {\footnotesize \text{$t/h$ (Hz)}} & {\footnotesize {$\kappa_xd^2/t\, (10^{-3})$}} & {\footnotesize {$\kappa_yd^2/t\,(10^{-3})$}} \\
\midrule
2.3(1) & 4.0(1) & 398(6) & 170(4) & 10(1) & 13(1) \\
7.5(4) & 7.1(1) & 582(9) & 78(3) & 26(1) & 35(1) \\
10.4(6) & 8.0(2) & 634(9) & 61(2) & 38(1) & 50(1) \\
33(2) & 12.0(2) & 816(11) & 25(1) & 135(1) & 174(1) \\
\end{tabular}
\end{ruledtabular}
\caption{\label{tab:hubbardparams}
Hubbard parameters for our system.
$V$ is the lattice depth along one direction (both lattices have the same depth).
For $U/t = 2.3(1)$, the next-nearest-neighbor hopping is $h \cdot \SI{12}{Hz}$.
}
\end{table}
%

\begin{figure}[t]
  \includegraphics[width=\columnwidth]{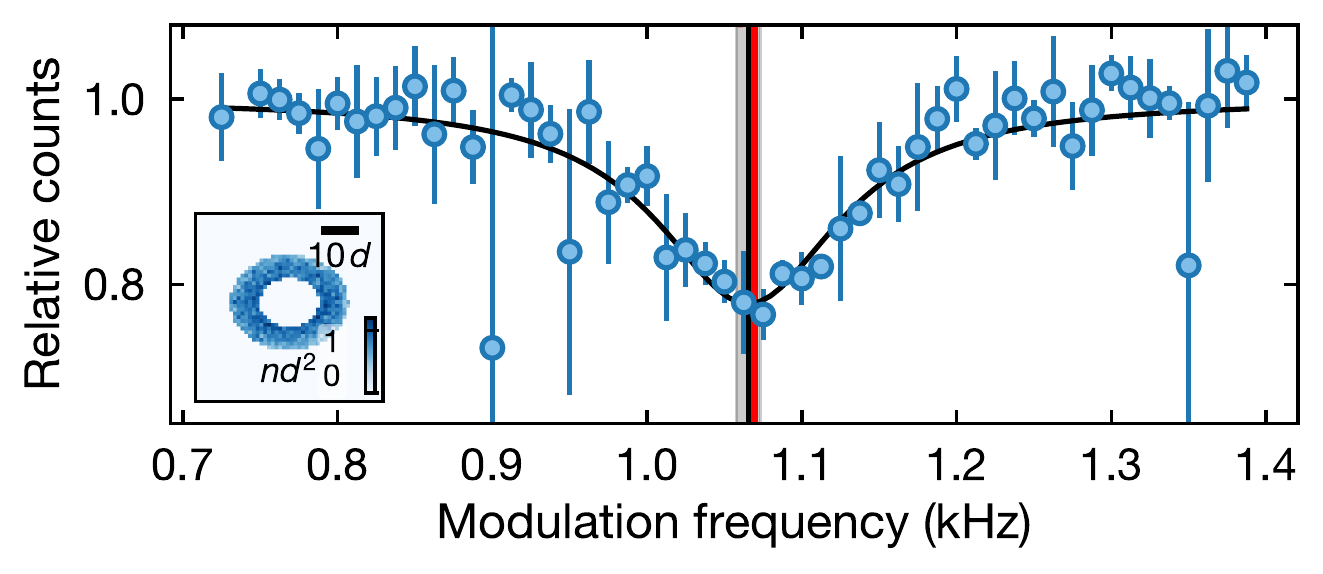}
  \caption{\label{fig:Ugg} 
  Direct measurement of $U$ with modulation spectroscopy for $U/t \approx \num{178}$. 
  Red line: expected value of $U$ according to the Wannier overlap.
  Black line: fit of a Lorentzian function.
  To enhance the signal-to-noise ratio, data have been evaluated in an elliptical shell where the cloud is mainly in the insulating phase (inset).
  }
\end{figure}

We calibrate the lattice depths along the two directions by measuring the band gap between the lowest and the second excited band with parametric modulation.
$U$ and $t$ are obtained by numerically solving the band structure for the measured band gap and $U$ is calculated as the Wannier overlap in the lowest band~\cite{jakschColdBosonicAtoms1998}.
The values of the relevant Hubbard parameters for this work are reported in Tab.~\ref{tab:hubbardparams}.
We cross-check the calculation by directly measuring $U$ with modulation spectroscopy for $V = 13\,\Erec$ ($U/t \approx 43$) and $19\,\Erec$ ($U/t \approx 178$).
For this measurement, we first tune the lattice depth to the desired value. 
We then modulate the depth of both lattices with an amplitude of \SIrange{2}{6}{\percent} and at the same time apply the pair removal pulse.
For modulation frequencies resonant with $U$, atom losses are enhanced.
For $13\,\Erec$, we measure $U = h \cdot \SI{886(8)}{Hz}$ for an expected value of $h \cdot \SI{856(8)}{Hz}$.
For $19\,\Erec$, we measure $U = h \cdot \SI{1065(8)}{Hz}$ for an expected value of $h \cdot \SI{1069(9)}{Hz}$ (Fig.~\ref{fig:Ugg}).

\subsection{Pair removal}
\label{sec:pair_removal}

\begin{figure}[t]
  \includegraphics[width=\columnwidth]{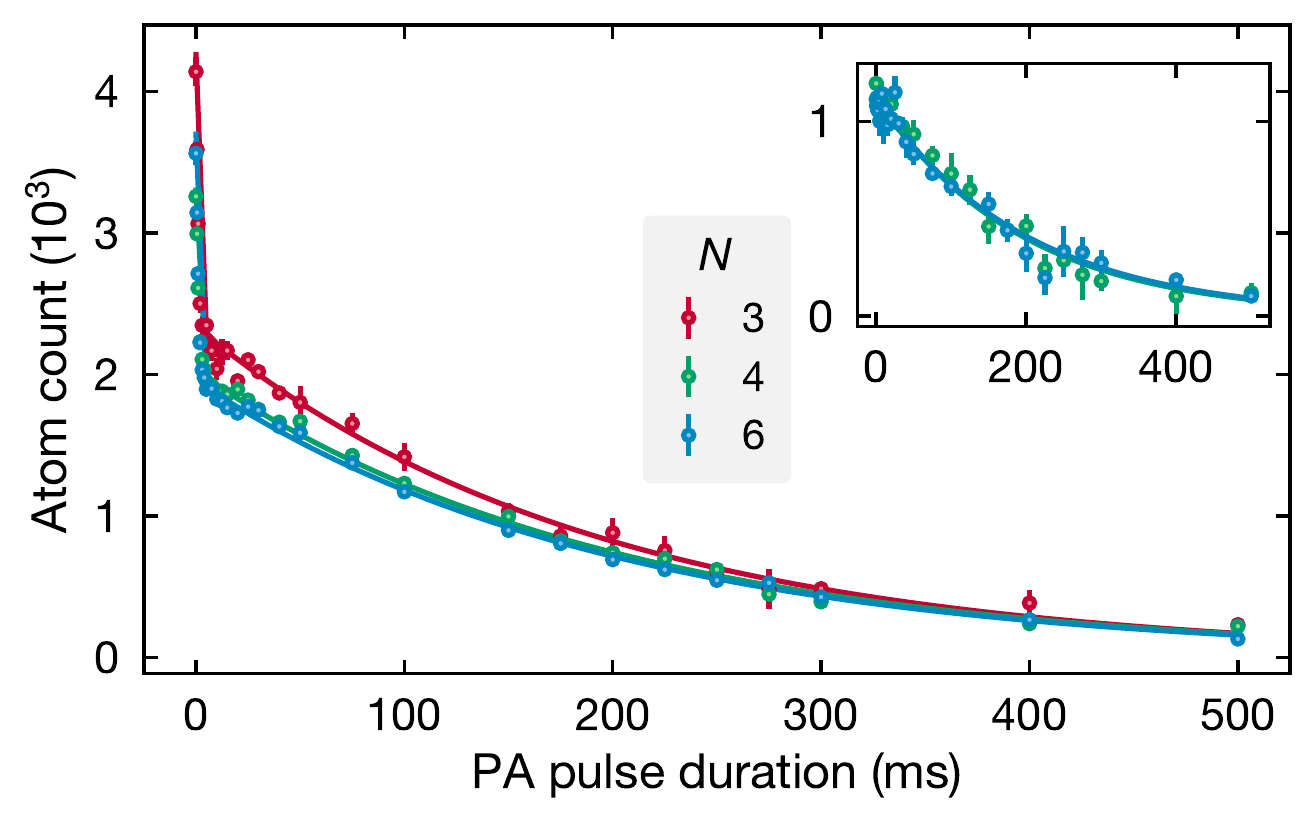}
  \caption{\label{fig:pa_efficiency} 
  Calibration of the photoassociation beam.
  Atom number as a function of the pulse duration for $\N = 3$ (red), $\N = 4$ (green) and $\N = 6$ (blue).
  We first tune the lattice depth to $12\,\Erec$ [$U/t = 33(2)$] and then quench it to $30\,\Erec$ before applying the photoassociation pulse.
  Inset: Same experiment repeated for a smaller atom number, such that $nd^2 \simeq 1$ in the center.
  }
\end{figure}

The pair-removal operation is performed with a photoassociation beam working on the $\sS{0} \rightarrow \tP{1}$ intercombination line, resonant with a molecular transition detuned by $-599.28(8)\,\MHz$ with respect to the single-particle transition with a bias magnetic field of \SI{1}{G}.
The beam is coplanar with the lattices and has a power of \SI{30}{mW}.
For pair removal, we first ramp the lattice to the desired $U/t$ value according to the ramps described in Sec.~\ref{sec:lattice_loading}, quench the depth to $30\,\Erec$ in \SI{0.5}{ms} and then apply the photoassociation pulse for \SI{10}{ms}. 

The on-site atom number after pair removal is:
\begin{equation}
  n_\text{PR} = \sum_{\alpha=1}^N \left(\alpha\bmod2\right) p_\alpha,
  \label{eq:PA}
\end{equation}
where $p_\alpha$ is the probability of having $\alpha$ particles on a given site ($\alpha p_\alpha$ is the average number of particles on a site with occupation $\alpha$ and $n = \sum_{\alpha=0}^N \alpha p_\alpha$) and we neglect the tunneling during the quench.
We correct Eq.~\ref{eq:PA} to take into account the experimental imperfections of the photoassociation beam:
\begin{multline}
  \tilde n_\text{PR} \simeq e^{-\gamma_s t}\left[ \sum_\alpha{\left({\alpha\bmod 2}\right)p_\alpha} \right.\\
 + \left. e^{-\gamma_d t}\left( \sum_\alpha{{2\, \lfloor \alpha/2 \rfloor} p_\alpha}\right)\right],
  \label{eq:PA_corrected}
\end{multline}
where $\lfloor\cdot\rfloor$ represents the floor function, $\gamma_s$ and $\gamma_s + \gamma_d$ represent the decay rate of the singlons and the doublons and we neglect the fast decay of the states with more than two particles per site~\cite{hofrichterDirectProbingMott2016}.
For ${\gamma_s, \gamma_d \to 0}$ we recover $\tilde n_\text{PR} \to n_\text{PR}$.

\subsubsection*{Efficiency calibration}

We calibrate the efficiency of the pair removal by looking at the atom losses as a function of the pulse duration in the deep Mott insulating regime [$U/t = 33(2)$]. 
Neglecting the sites occupied by more than two particles, we fit a double-exponential model:
\begin{equation}
N_p = e^{-\gamma_s t} \left( N_s+ N_d e^{-\gamma_d t}\right),
\end{equation}
where $N_p = N_s + N_d$ is the total atom number, $N_s$ the number of singlons, $N_d$ the number of doublons.
We obtain $1/\gamma_d = 1.2(2)\,\ms$ and $1/\gamma_s = 200(11)\,\ms$, which we find to be independent of \N inside the uncertainties (see Fig.~\ref{fig:pa_efficiency}).
We cross-checked the value of $\gamma_s$ by repeating the same experiment in a small Mott insulator with $\approx 1\times 10^3$ atoms and no doublons (inset of Fig.~\ref{fig:pa_efficiency}). 
For a pulse duration of \SI{10}{ms}, we remove all the doublons within our detection sensitivity and about $5\,\%$ of the singlons.
We take this into account when calculating the theory curves in Fig.~2.

\subsection{Characterization of the potential}

\subsubsection{Trap frequencies}
\label{sec:trap_freqs}

\begin{figure}[t]
  \includegraphics[width=\columnwidth]{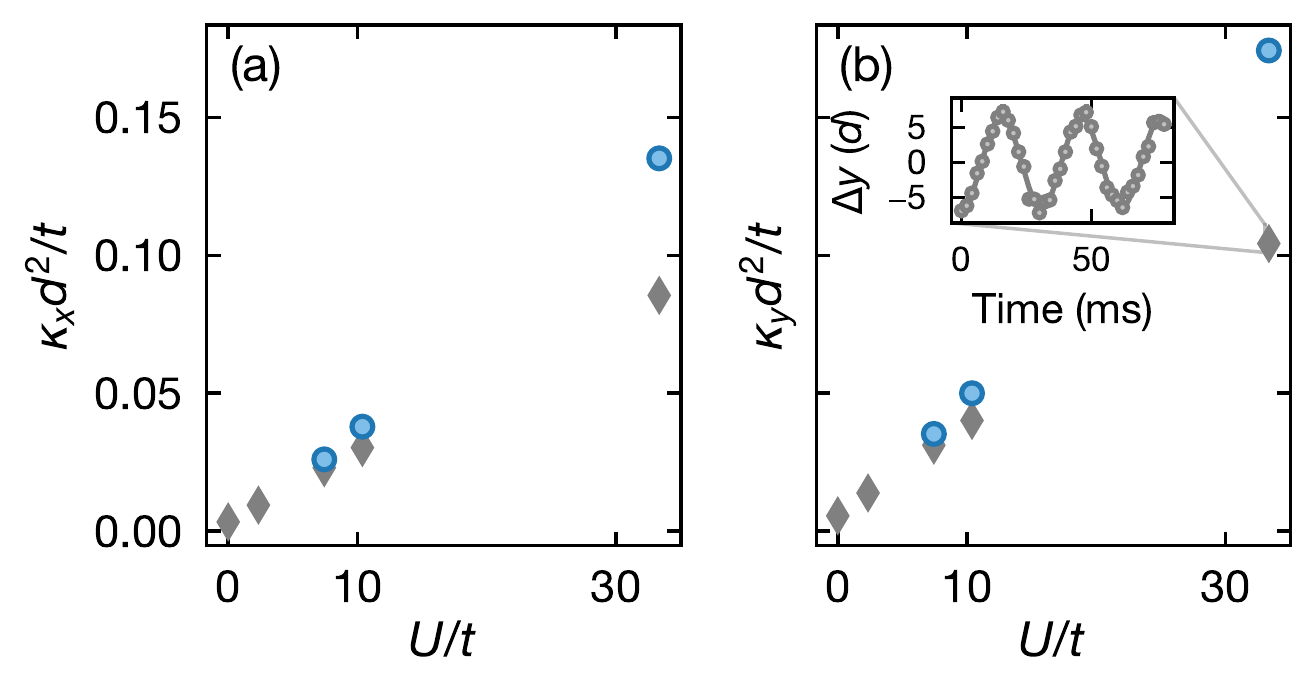}
  \caption{\label{fig:trap_frequencies} 
  Coefficients of the harmonic confinement along the lattice axes $(x,y)$. 
  Circles correspond to the harmonic confinement according to the EoS fit described in the text.
  Diamonds correspond to an independent calibration measuring the frequencies of the COM oscillations around the center of the trap.
  Inset: example of COM oscillations for $U/t = 33(2)$.
  }
\end{figure}

Across the region sampled by the atomic cloud, the confinement due to the Gaussian lattice beams is approximately harmonic, such that $\mu(x,y) = \mu_0 - \frac{1}{2}\left(\kappa_x x^2 + \kappa_y y^2\right)$.
The two coefficients are determined by a fit of the EoS for each lattice depth with free parameters $T/t, \mu_0/t, \kappa_xd^2/t, \kappa_yd^2/t$.
We verify that the fourth order anharmonic correction coming from the Gaussian profile, calculated for the measured beam waist and power of the beams, is negligible for the size of the cloud.
However, the values obtained for $\kappa_x$ and $\kappa_y$ do not fully agree with the independent measurement of the frequency of the oscillatory motion of the atoms in the combined potential.
More specifically, we load a spin-polarized cloud in the combined potential of the vertical lattice and one in-plane lattice at a time.
We measure the center of mass (COM) oscillation after an initial displacement of about $5\, d$ along the lattice direction and we verify that the Rayleigh range contribution to the combined potential is negligible.
A comparison between the two sets of values can be found in Fig.~\ref{fig:trap_frequencies}.

\subsubsection{Fit comparisons}

\begin{figure}[t]
  \includegraphics[width=\columnwidth]{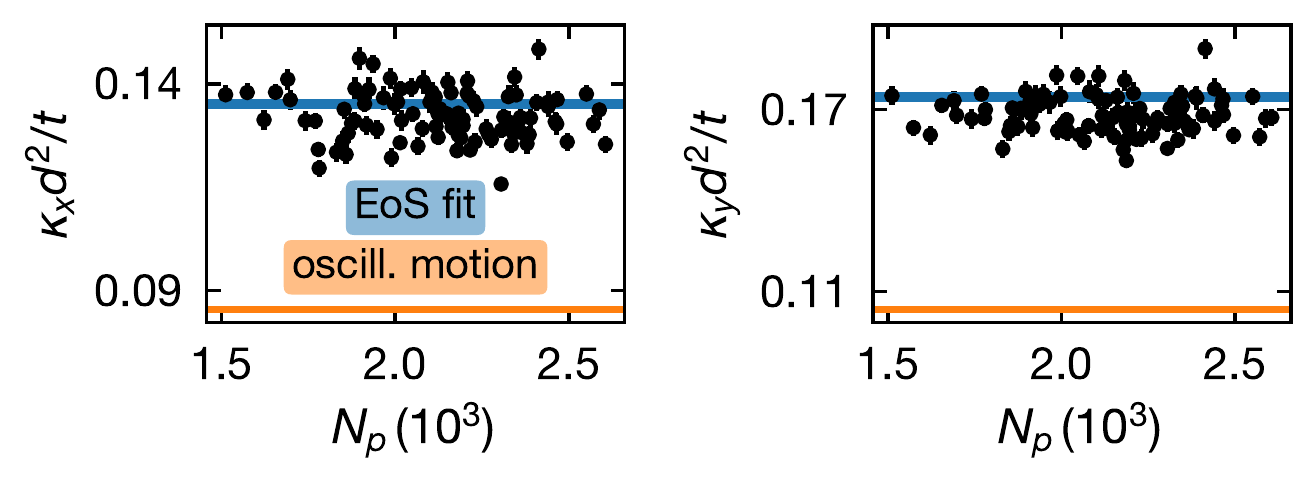}
  \caption{\label{fig:radial_dep} 
  Dependency of $\kappa_x$ and $\kappa_y$ with the atom number $N_p$.
  Each point corresponds to a single realization with $\N=6$ and $U/t=33(2)$.
  For each realization we fit ($T/t$, $\kappa_xd^2/t$, $\kappa_yd^2/t$, $\mu_0/t$) with NLCE.
  Blue lines: values used in the main text for this $U/t$ ratio.
  Orange lines: values determined with the independent ``oscillatory motion" calibration.
  }
\end{figure}
%

\begin{figure}[t]
  \includegraphics[width=\columnwidth]{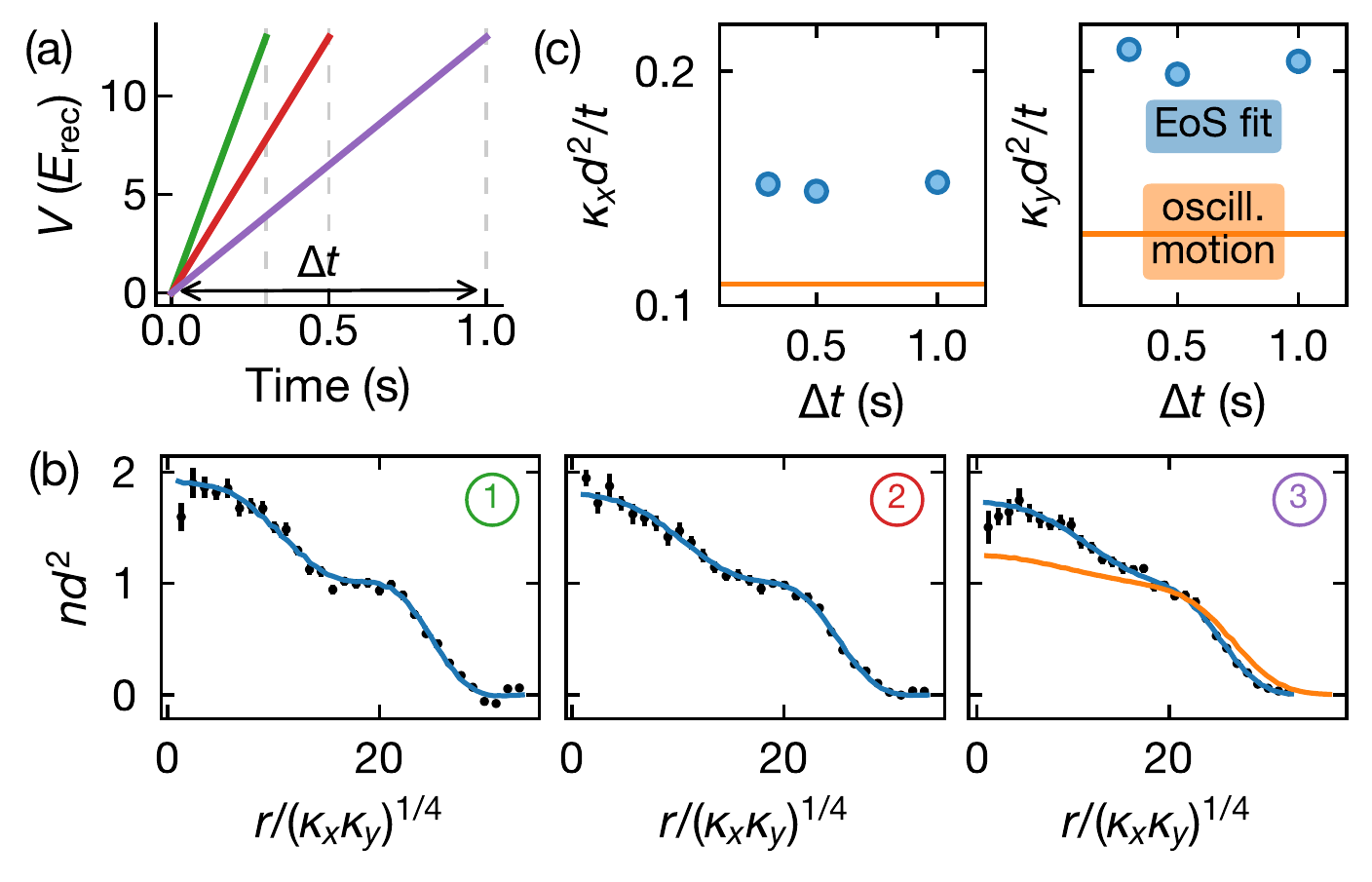}
  \caption{\label{fig:scan_ramp_time}
  Robustness of the EoS fit as a function of the ramp duration.
  (a) We tune a SU(6) sample to $U/t \approx 44$ with a linear ramp of duration $\Delta t = \SI{0.3}{s}$ (green), \SI{0.5}{s} (red), \SI{1}{s} (purple).
  (b) For each ramp duration $\Delta t$, we fit the density profile assuming an harmonic potential with coefficients determined by the ``oscillatory-motion" calibration (orange) and by leaving the coefficients free in the EoS fit (blue).
  Here, we plot the results of the fit with respect to the spacial radial profile.
  For the first two frames, the ``oscillatory-motion" fit fails.
  (c) Comparison of the harmonic potential coefficients. 
  }
\end{figure}

A possible explanation for the discrepancy in the shape of the inferred potential described in Sec.~\ref{sec:trap_freqs} might be a lack of adiabaticity and equilibration during the tuning of $U/t$.
However, the temperature measured with the fluctuation-dissipation theorem matches the one obtained from the fit of the EoS.
This is not the case if we use the chemical potential coming from the oscillatory-motion calibration.
In particular, we verify that for this potential,  $\N=6$ and $U/t=33(2)$ we obtain a temperature inside the regime of convergence of our theory models when we use the FDT, but which does not quantitatively match the temperature returned by the fit of the EoS with the same potential.

Moreover, we verify that the values of $\kappa_x$ and $\kappa_y$ determined by the EoS fit are robust against atom number variation (see Fig.~\ref{fig:radial_dep}).

Finally, we observe that $\kappa_x$ and $\kappa_y$ are also insensitive to the lattice ramp time duration.
In Fig.~\ref{fig:scan_ramp_time} we show a test which consists of tuning the lattice depth of an SU(6) sample from zero to $13\, \Erec$ ($U/t \approx 44$) with a linear ramp of different durations $\Delta t$ between \SI{300}{ms} and \SI{1}{s}.
For each $\Delta t$, we fit the EoS with second-order high-temperature series expansion (HTSE-2) and leave ($T/t$, $\kappa_xd^2/t$, $\kappa_yd^2/t$, $\mu_0/t$) as free fit parameters.
We observe that the values of $\kappa_x$ and $\kappa_y$ returned by the fit for different $\Delta t$ are compatible among each other [$\kappa_xd^2/t = 0.151(2)$, $\kappa_yd^2/t = 0.204(3)$].
However, they are incompatible with the ones predicted by the independent ``oscillatory-motion" calibration described in Sec.~\ref{sec:trap_freqs} ($\kappa_x^\text{osc}d^2/t \simeq 0.109$, $\kappa_y^\text{osc}d^2/t \simeq 0.131$).
If we use these values for the fit, it fails for $\Delta t = \SI{0.3}{s}$ and \SI{0.5}{s} and it returns high residuals for $\Delta t = \SI{1}{s}$, failing to reproduce the cloud shape especially in the center [Fig.~\subfigref{scan_ramp_time}{c}].
We conclude that if the mismatch between the two different harmonic potential parameters comes from a lack of equilibration during the tuning of the lattice depth, the time scale to achieve this equilibration significantly exceeds the experimentally accessible timescales.

\subsubsection{Anharmonicities}
\label{sec:potentialmapping}

We characterize the anharmonic corrections of the potential by an ensemble with a high temperature $T\gg t$ into the lattice and applying the atomic limit model with $t=0$.
In this way, we can generate a spatial map of the chemical potential and observe that the functional modeling as a harmonic trap is a good approximation.
In particular, we estimate the anharmonic corrections being less than $0.03\,\mu_0$ over the whole region of interest in the deep Mott insulating regime.

\subsection{Imaging techniques}
\label{sec:imaging}

\begin{figure}[t]
  \includegraphics[width=\columnwidth]{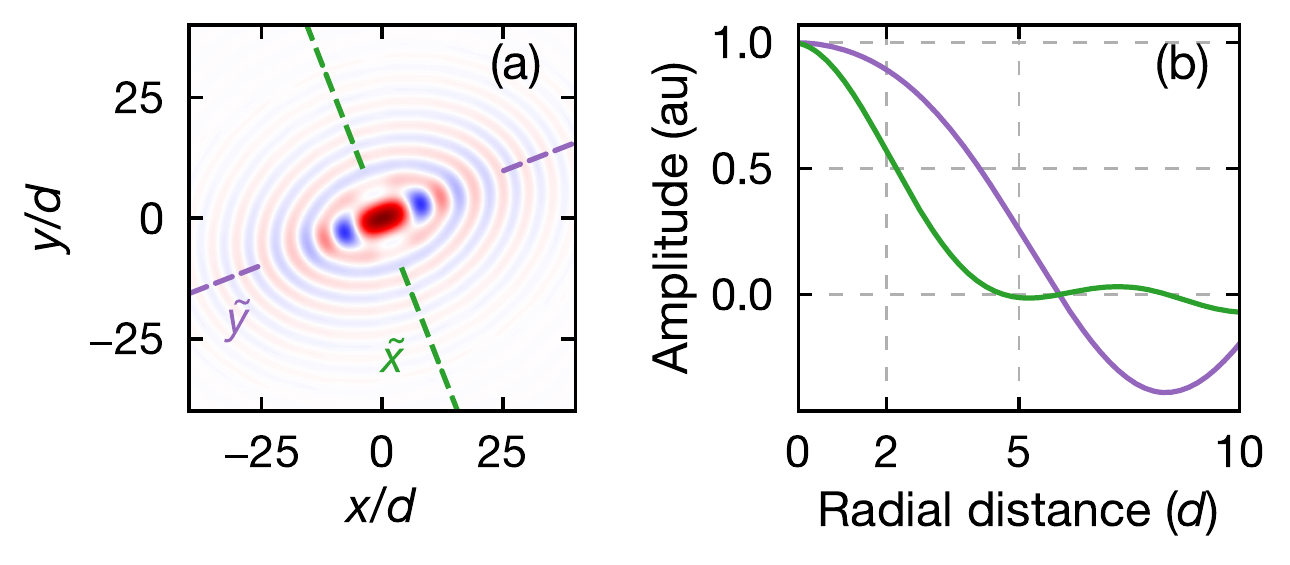}
  \caption{\label{fig:psf} 
  (a) Reconstructed point spread function (PSF).
  $x$ and $y$ have been labeled according to the convention defined in Fig.~1.
  The color scale is in arbitrary units and varies between -1 (blue) and 1 (red).
  (b) Cuts along the main axes of the PSF [dashed lines in (a)].
  }
\end{figure}

We use saturated in situ absorption imaging on the stretched $\sS{0} \rightarrow \sP{1}$ transition with circularly polarized light in the presence of a small magnetic field offset $\simeq \SI{1}{G}$ and a pulse duration of \SI{5}{\micro s}.
We determine the density with the modified Lambert-Beer law~\cite{reinaudiStrongSaturationAbsorption2007}, which accounts for the high-intensity effects:
\begin{equation}
  n(x, y) = \frac{1}{\sigma}\left[ \log{\left(\frac{I_\text{in}}{I_\text{out}}\right)} +  \frac{I_\text{in} - I_\text{out}}{I_\text{sat}^\text{eff}}\right],
  \label{eq:imaging}
\end{equation}
where $n(x, y)$ is the density at pixel position $(x, y)$ and $I_\text{in} = I_\text{in}(x,y)$ and $I_\text{out}=I_\text{out}(x,y)$ are respectively the incident light and the light after absorption.
We calibrate the effective saturation intensity $I_\text{sat}^\text{eff}$ by varying $I_\text{in}/I_\text{sat}$ between 2 and 8 (where $I_\text{sat} = \pi h c \Gamma /(3 \lambda^3) \simeq \SI{60}{mW/cm^2}$ is the saturation intensity of the transition with wavelength $\lambda$ and linewidth $\Gamma$) and minimizing the variation of the density profile as described in Ref.~\cite{reinaudiStrongSaturationAbsorption2007}.
We find $I_\text{sat}^\text{eff}/I_\text{sat} = 3.0(2)$ with variations of less than 5\% between spin mixtures.
We obtain compatible values in the 3D dipole trap, in the 2D bulk and in the deep Mott insulating regime with variations smaller than 5\%.
The effective cross section $\sigma$ is calibrated for each spin mixture by determining the minimum of the compressibility $\kappa = \partial n/\partial \mu$ near the insulating regime at $nd^2 \simeq 1$ (see also Fig.~\ref{fig:compressibility}).
We obtain $\sigma/\sigma_0 = [0.310(3), 0.320(3), 0.321(3)]$ respectively for $\N= [3, 4, 6]$, where $\sigma_0 = 3\lambda^2/(2\pi)$ is the photon resonant cross section.
As an independent measurement, we extract the effective cross section from the density shot noise of a thermal sample in the 2D bulk according to the method described in Ref.~\cite{hungSituImagingAtomic2014}. 
In particular we make use of the fact that $\langle | \delta n_0(\textbf{k})|^2\rangle$ is proportional to $\sigma N_p\mathcal{M}^2(\textbf{k})$,
where $\delta n_0$ are the measured local density fluctuations per pixel, $\textbf{k}$ is a vector in Fourier space, $N_p$ is the total atom number and $\mathcal{M}(\textbf{k})$ is the modulation transfer function of the imaging system.
By modeling $\mathcal{M}$ with aperture, intensity attenuation and aberrations as free fit parameters, we determine both $\sigma$ and $\mathcal{M}$ \cite{hungSituImagingAtomic2014}.
We obtain $\sigma/\sigma_0 = [0.35(1), 0.38(1), 0.384(7)]$ respectively for $\N = [3, 4, 6]$.
We attribute the differences between the values to cooperative optical response effects at high densities~\cite{chomaz2012,ruiSubradiantOpticalMirror2020}.

\subsubsection*{PSF modeling}

The point spread function (PSF) as reconstructed from $\mathcal{M}$ \cite{hungSituImagingAtomic2014} is shown in Fig.~\ref{fig:psf}. 
The imaging imperfections are taken into account by convolving the 2D profile of the density predicted by the theory with the PSF.
We estimate the systematic error in the determination of the EoS by varying the HWHM of the reconstructed PSF.
Testing the sensitivity of the fit parameter results, we find that, in the deep Mott insulating regime, where the effects of the PSF are most relevant, a variation of \SI{50}{\percent} in the size of the HWHM causes only a change of \SI{4}{\percent} in the entropy [for $U/t = 33(2)$ and $\N =6$].

\subsection{Fit method and parameters for the EoS}

We fit a 2D model of the density with fixed $\kappa_x$, $\kappa_y$ and cross section determined with a separate fit as described in Sec.~\ref{sec:trap_freqs} and~\ref{sec:imaging}.
Due to small position fluctuations during imaging, we perform an initial fit of each image to determine the center of the trap.
We use this to align the centers of the distributions and perform a fit of the average of several realizations with similar atom number (15 realizations for the dataset of Fig.~2).

\begin{table}[t]
\begin{ruledtabular}
\begin{tabular}{lSlllll}
\N & $U/t$ & $N_p$ ($\times 10^3$) & Method & $T/t$ & $T/U$ & $s/\kB$ \\
\midrule
3 & 2.3(1) & 1.98(2) & DQMC & 0.40(4) & 0.17(2) &  \\
  & 7.5(4) & 1.96(2) & NLCE-6 & 1.24(3) & 0.167(4) & 1.50(6) \\
  & & & NLCE-7 & 1.24(3) & 0.167(4) & 1.50(6) \\
  & 10.4(6) & 1.94(3) & NLCE-6 & 1.69(4) & 0.163(4) & 1.55(6) \\
  & & & NLCE-7 & 1.69(4) & 0.163(4) & 1.55(6) \\
  & & & HTSE-2 & 1.69(4) & 0.163(4) & 1.56(6) \\
  & 33(2) & 1.99(2) & NLCE-6 & 5.0(1) & 0.149(3) & 1.51(4) \\
  & & & NLCE-7 & 5.0(1) & 0.149(3) & 1.51(4) \\
  & & & HTSE-2 & 5.0(1) & 0.149(3) & 1.51(4) \\
\midrule
4 & 2.3(1) & 1.99(1) & DQMC & 0.32(3) & 0.13(1) &  \\
  & 7.5(4) & 1.99(1) & DQMC & 0.97(3) & 0.130(4) & 1.61(7) \\
  & & & NLCE-4 & 0.98(3) & 0.131(3) & 1.59(6) \\
  & & & NLCE-5 & 0.94(3) & 0.126(4) & 1.57(7) \\
  & 10.4(6) & 1.99(1) & DQMC & 1.36(4) & 0.131(4) & 1.73(6) \\
  & & & NLCE-4 & 1.42(4) & 0.137(4) & 1.73(6) \\
  & & & NLCE-5 & 1.41(4) & 0.136(4) & 1.72(6) \\
  & & & HTSE-2 & 1.43(4) & 0.138(3) & 1.76(7) \\
  & 33(2) & 1.99(1) & NLCE-4 & 4.62(9) & 0.139(3) & 1.78(5) \\
  & & & NLCE-5 & 4.62(9) & 0.139(3) & 1.78(5) \\
  & & & HTSE-2 & 4.63(9) & 0.139(3) & 1.78(5) \\
\midrule
6 & 2.3(1) & 1.99(1) & DQMC & 0.30(1) & 0.127(1) &  \\
  & 7.5(4) & 1.99(1) & DQMC & 0.91(3) & 0.122(4) & 1.95(9) \\
  & & & NLCE-3 & 0.94(3) & 0.126(4) & 1.97(8) \\
  & & & NLCE-4 & 0.91(2) & 0.122(3) & 1.90(7) \\
  & 10.4(6) & 2.05(1) & DQMC & 1.35(4) & 0.131(4) & 2.11(8) \\
  & & & NLCE-3 & 1.40(4) & 0.135(3) & 2.10(8) \\
  & & & NLCE-4 & 1.38(4) & 0.133(4) & 2.08(8) \\
  & & & HTSE-2 & 1.48(3) & 0.142(3) & 2.15(9) \\
  & 33(2) & 2.00(1) & NLCE-3 & 3.99(8) & 0.120(2) & 2.12(6) \\
  & & & NLCE-4 & 3.99(8) & 0.120(2) & 2.12(6) \\
  & & & HTSE-2 & 4.06(8) & 0.122(2) & 2.13(6) \\
\end{tabular}
\end{ruledtabular}
\caption{\label{tab:fig23}
Fit parameters for the EoS shown in Fig.~2 and comparison between different methods.
For each sample, we consider 15 shots postselected according to the total atom number.
For HTSE and NLCE methods, the number indicates the order.
The agreement between two consecutive orders indicates that NLCE has converged.
In Fig.~2 we plot the highest NLCE order if available, DQMC otherwise.
The uncertainties correspond to the values returned by the fit and do not take into account additional systematic errors.
}
\end{table}

In Tab.~\ref{tab:fig23} we report the values returned by the EoS fit shown in Fig.~2 and a comparison between different numerical methods.

The fit results do not take into account the uncertainties on $U/t$.
By fitting different values of $U/t$ to the same data, we estimate the systematic error on the entropy per particle and the temperature to be about $1\%$ and $0.015\,U$ respectively.

\subsection{Equation of state for higher temperatures}

\begin{figure}[t]
  \includegraphics[width=\columnwidth]{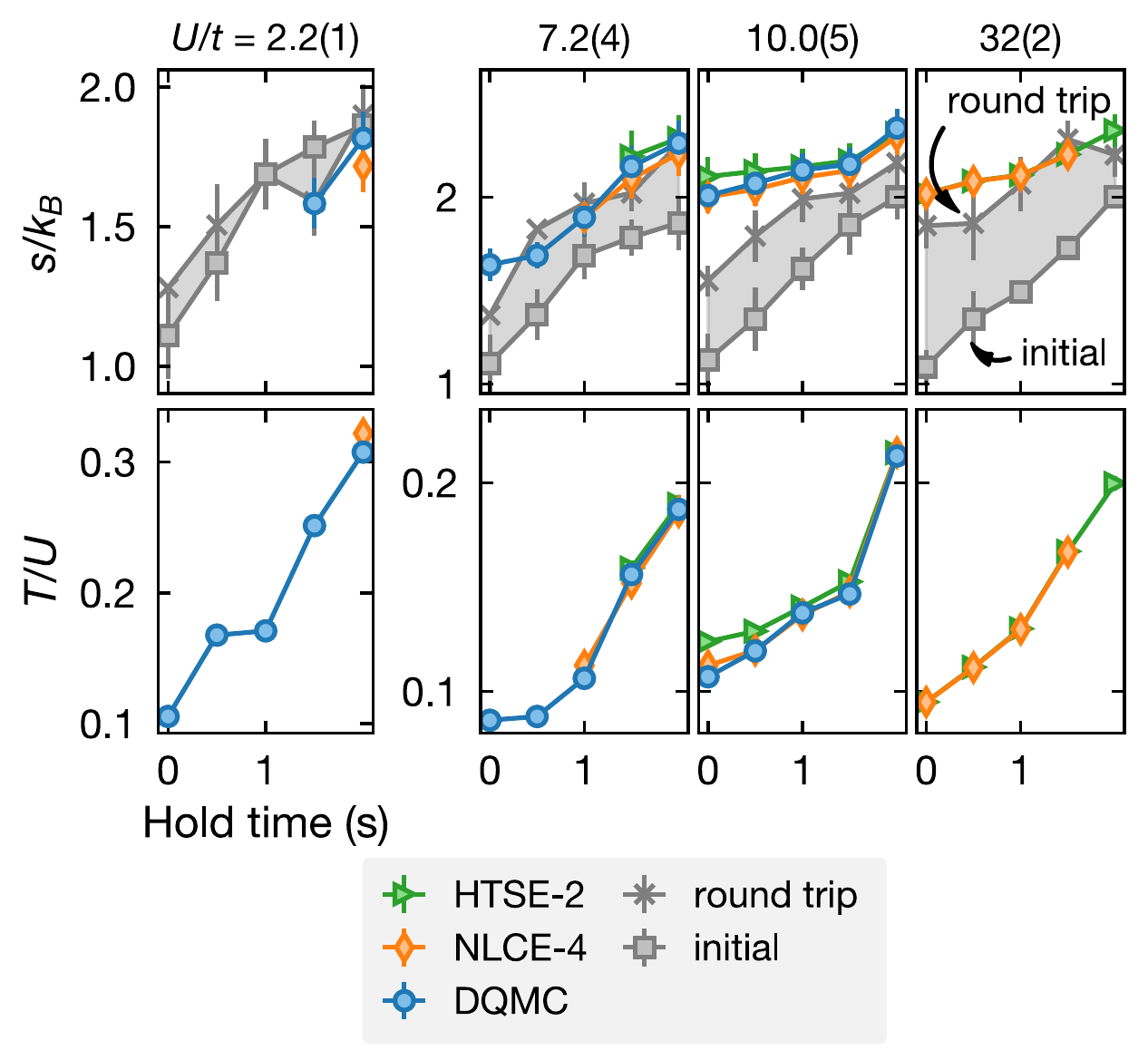}
  \caption{\label{fig:scan_temp_su6} 
  Equation of state for high temperature for $\N = 6$.
  By varying the hold time in the 2D bulk before ramping up the lattices, we increase the initial entropy per particle (grey squares).
  After loading into the lattice, we perform a fit of the density obtained by the average of 15 frames with $N_p \approx 2.2\times 10^3$ and a s.e.m. of 15 to 35 for each combination of $U/t$ and hold time, after atom number postselection and COM alignment.
  We fit the same datapoint with different methods and compare the results.
  We fit DQMC (blue circles), NLCE of order 4 (orange diamonds) and HTSE of order 2 (green triangles).
  After loading into the lattice we perform a round-trip experiment by symmetrically ramping down the lattices and measuring the entropy in the bulk (grey crosses).
  For this measurement, the vertical bandgap was $h \times \SI{3.63(1)}{kHz}$ and the confinement's parameters $\kappa_x$ and $\kappa_y$ have been calibrated according to the same method presented in the main text (for $\N = 6$ only).
  }
\end{figure}

In Fig.~\ref{fig:scan_temp_su6} we probe the equation of state as a function of the initial entropy for $\N=6$.
We increase the entropy by holding the atoms in the 2D bulk before the lattice ramps.
In this way, we can vary the entropy per particle $s$ between approximately $1\,\kB$ and $2\,\kB$ (grey squares).
After loading into the lattices, we fit the density and compare different numerical methods, including DQMC, NLCE and HTSE.
For high initial entropy and large interactions, the methods agree very well with each other.
For lower initial entropy, first HTSE and then NLCE begin to deviate.
We also perform a round-trip experiment as described in the main text (grey crosses).
For large initial entropies, we observe a good agreement between the entropies in the lattice and after the round trip experiment, indicating that the increase is mainly occurring during the ramp up.

\subsection{Equation of state in the bulk}
\label{sec:bulkEoS}
The entropy per particle in 2D prior to turning on the in-plane lattices is calculated from a fit of the equation of state for a weakly-interacting Fermi gas. 
In particular, we obtain the temperature from the fit of the in situ density according to:
\begin{equation}
n(\mu, T) = -\frac{\partial \Omega}{\partial \mu} = -N\frac{m}{2\pi\hbar^2\beta} \text{Li}_1(-e^{\beta \mu}),
\label{eq:density}
\end{equation}
where $\Omega$ is the grand potential, $\text{Li}_1$ is the polylogarithm of order 1, $\beta=1/(\kB T)$, and $m$ is the mass of one \yb atom.
The harmonic potential is taken into account in local density approximation.
The effect of interactions is taken into account by introducing a correction to the chemical potential~\cite{engelbrechtLandauFunctionDilute1992}:
  \begin{multline}
  \mu_\text{int}(\beta) \approx \frac{1}{\beta}\left[1 + 2g(\N-1) \right.\\
  \left.+ 4g^2(\N-1)(1-\log{2})\right] \log{\left(e^{\beta \EF}-1\right)},
    \label{eq:chem_pot_int_finite_T}
  \end{multline}
  where \EF is the Fermi energy and $g$ is the interaction parameter:
\begin{equation}
  g(n, N) = \frac{1}{\log{2}-2\log{\left(k_\text{F} a_\text{2D}\right)}}.
\end{equation}
Here, $\kF = \sqrt{4\pi n /N}$ is the local Fermi vector and $a_\text{2D} \simeq \SI{2.5e-3}{a_0}$ is the 2D scattering length.
The density-dependency of the interaction parameter is evaluated self-consistently in the fit routine similarly to what has been done in 3D in Ref.~\cite{sonderhouseThermodynamicsDeeplyDegenerate2020}.
We convolve the predicted density profile with the PSF to take into account imaging imperfections.
The entropy is calculated from the temperature returned by the fit and the spectrum of the 3D non-isotropic harmonic oscillator~\cite{kohlThermometryFermionicAtoms2006}.
We determine a correction to the entropy due to the interactions between \SI{5}{\percent} and \SI{20}{\percent} compared to the one returned by the fit of a non-interacting Fermi profile.

\subsection{Fluctuations}

\subsubsection{Binning, PSF and shot noise}
\label{sec:binning}

\begin{figure}[t]
  \includegraphics[width=\columnwidth]{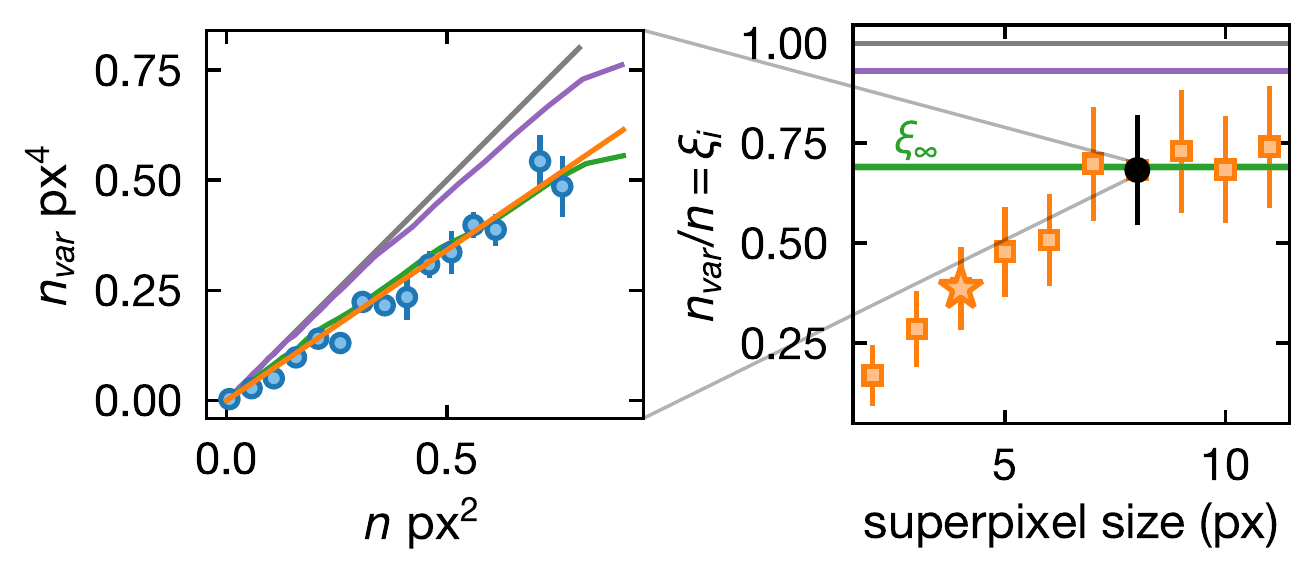}
  \caption{\label{fig:thermal_slope} 
  Fluctuations in the bulk for $\N = 6$.
  Left: density variance calculated with superpixels of size $\SI{8}{px}\times\SI{8}{px}$.
  Blue points: measured density variance.
  Grey line: for a thermal, non-interacting cloud $\text{var}(n)/n = 1$.
  Purple line: theoretical prediction estimated from the temperature returned by a non-interacting Fermi fit [$T/\TF = 1.37(1)$].
  Green line: theoretical prediction according to a weakly-interacting Fermi fit [$T/\TF = 1.02(1)$].
  Orange line: linear fit of the measured density fluctuations.
  Right: slope $n_\text{var}/n = \xi_i$ returned by the linear fit as a function of the superpixel size $i$ [($i \times i$) $\text{px}^2$ is the superpixel area].
  Star ($i=4$): superpixel size used in the lattice.
  }
\end{figure}

We spatially bin each frame in square ``superpixels'' and compute the chemical potential $\mu$ and the local density variance in each of them.

We take into account the PSF originating from the binning by measuring the density fluctuations for a SU(6) thermal cloud ($T \gg \TF$) in the bulk with the same binning.
For a non-interacting cloud  with point-like PSF and $T/\TF \gg 1$, $\text{var}(\hat n)/\langle \hat n \rangle \simeq 1$.
The binning lowers this ratio, which can be used as a calibration scaling factor for the fluctuations in the lattice.
However, we also take into account additional corrections due to the temperature and the interactions in the bulk.
We do so by fitting the cloud with a weakly-interacting model (see Sec.~\ref{sec:bulkEoS}) and determine $T/\TF = 1.02(1)$.
For this temperature and interaction strength, we expect $\text{var}(\hat n)/\langle \hat n \rangle \equiv \xi_\infty = 0.69(1)$ for small densities (see Fig.~\ref{fig:thermal_slope}).
The binning to finite-size superpixels introduces an additional correction $\xi_\infty \rightarrow \xi_i$, where $i$ is the linear size of the superpixel.
For $\SI{4}{px}\times \SI{4}{px}$ superpixels, the size that we use in Fig.~3, we measure the correction $\xi_4 = 0.39(1)$.

In the lattice, we first do the binning, then subtract the photon shot noise as the average density at the edge of the region of interest and finally rescale the variance by the factor $\left(\xi_4 \,\sigma^\text{bulk}/\sigma^\text{lat}\right)^{-1}$, which also takes into account the correction to the cross section described in Sec.~\ref{sec:imaging}.

\subsubsection{Compressibility}
\label{sec:compressibility}

\begin{figure}[t]
  \includegraphics[width=\columnwidth]{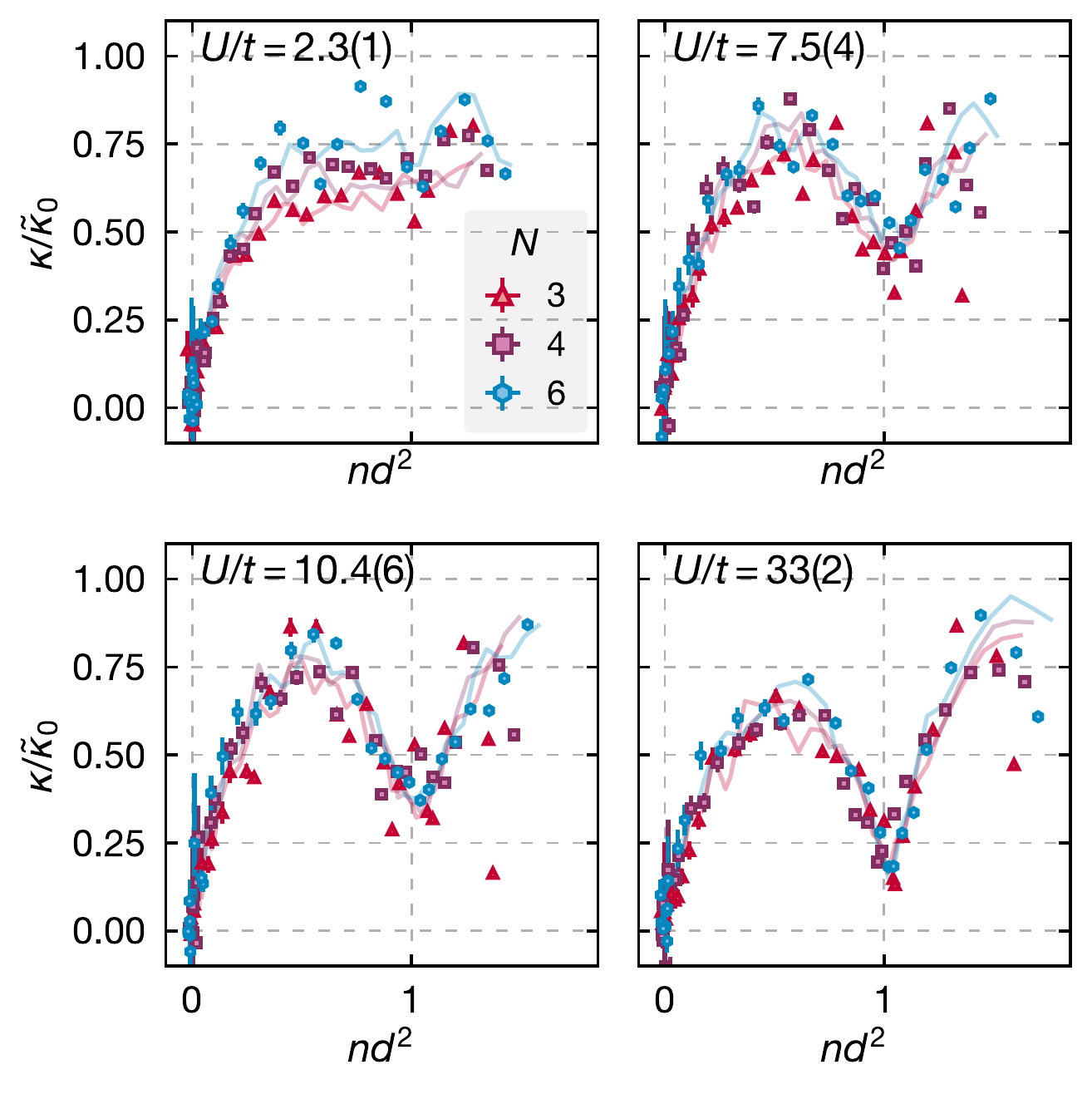}
  \caption{\label{fig:compressibility} 
  Isothermal compressibility $\kappa = \at{\partial n/\partial \mu}{T}$ for $\N = 3$ (red), $\N = 4$ (purple), and $\N = 6$ (blue).
  $\tilde \kappa_0$ is the compressibility of a SU(6) non-interacting gas at zero temperature.
  For each combination of \N and $U/t$, we numerically differentiate the density with respect to the chemical potential.
  The dataset is the same as Figs.~2 and 3.
  Points correspond to the differentiation of the experimental data and lines correspond to the differentiation of the theoretical curves.
  }
\end{figure}

To calculate the isothermal compressibility $\kappa = \at{\partial n/\partial \mu}{T}$, we first bin the experimental data in 1D as $n(\mu)$ and then numerically perform a 3-point differentiation.
Fig.~\ref{fig:compressibility} shows the compressibility for the $\N = 6$ dataset of Fig.~3 and a comparison with the datasets for $\N = 3$ and 4 shown in Fig.~2.

\subsubsection{FDT Thermometry}

\begin{figure}[t]
  \includegraphics[width=\columnwidth]{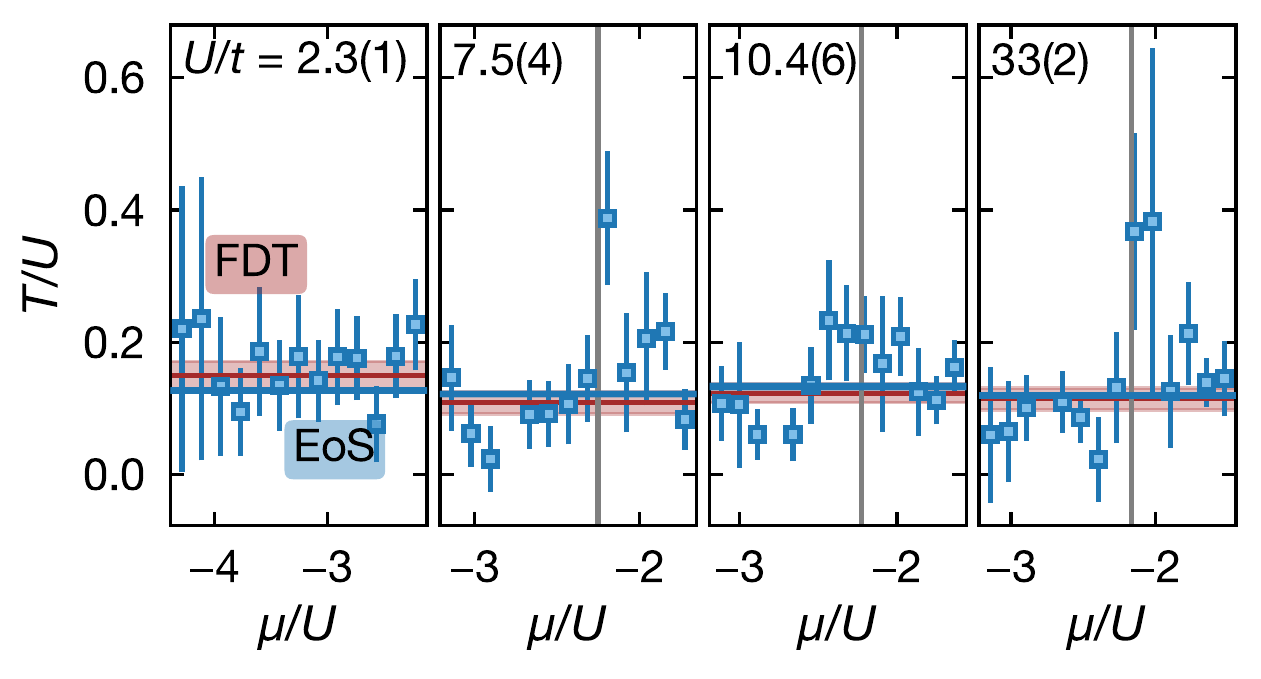}
  \caption{\label{fig:local_temp} 
  Local temperature for the $\N = 6$ dataset shown in Fig.~3.
  Blue line: temperature according to the fit of the EoS.
  Red line: temperature according to the fit of the FDT.
  Vertical line: $\mu (nd^2 =1)$.
  The local temperature for small and large densities is in general comparable, hinting at global thermal equilibrium.
  }
\end{figure}
%

\begin{figure}[t]
  \includegraphics[width=\columnwidth]{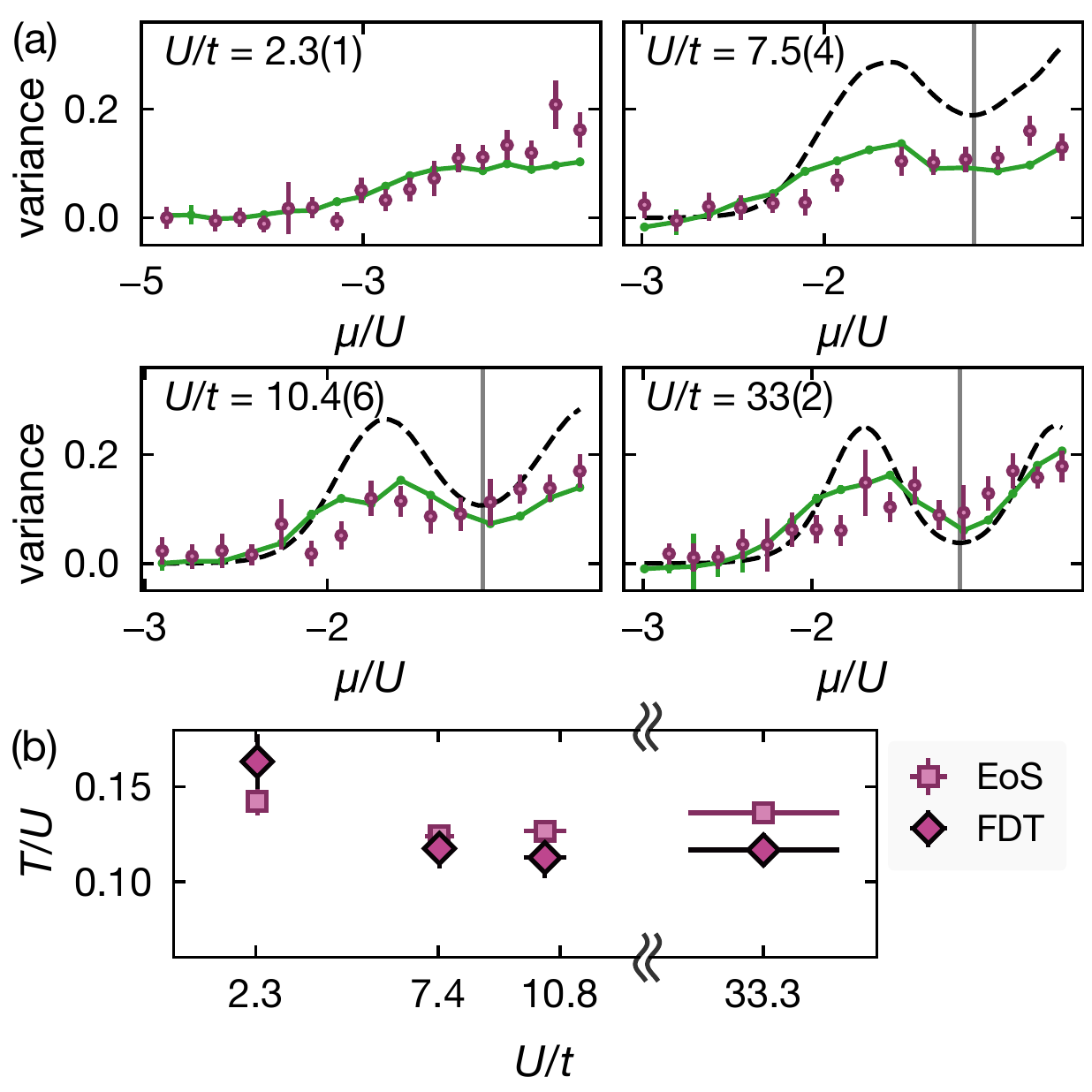}
  \caption{\label{fig:fluctuations4} 
  (a)~Measured density fluctuations (purple) for $\N=4$ as a function of the chemical potential for different interaction strengths.
  The data points have been obtained from the variance of 35 frames according to the same binning and evaluation method of Fig.~3.
  The green line corresponds to the numerically-differentiated compressibility $\kappa$ times the temperature $T_\text{EoS}$ obtained from the EoS-fit of the averaged data.
  The vertical line indicates $\mu (nd^2 =1)$.
  The dashed line corresponds to the on-site density fluctuations $\delta n_0^2 = \langle \hat n^2\rangle - \langle \hat n\rangle^2$ calculated with NLCE for $T_\text{EoS}$. 
  (b)~Comparison of the temperatures $T_\text{FDT}$ (dark purple diamonds) and $T_\text{EoS}$ (light purple squares).
  Error bars are the s.e.m.
  }
\end{figure}
%

\begin{figure}[t]
  \includegraphics[width=\columnwidth]{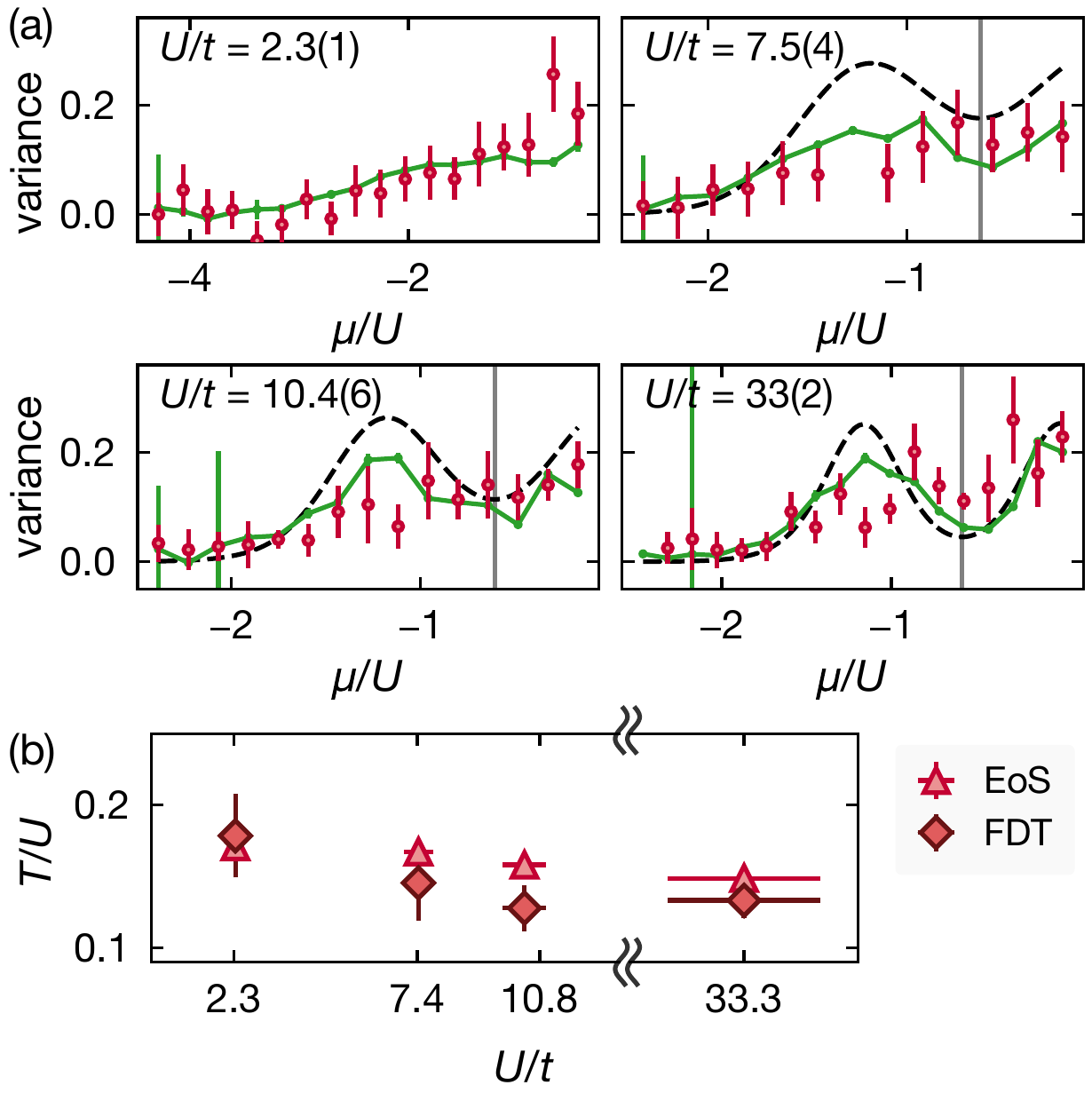}
  \caption{\label{fig:fluctuations3} 
  (a)~Measured density fluctuations (red) for $\N=3$ as a function of the chemical potential for different interaction strengths.
  The data points have been obtained from the variance of 15 frames according to the same binning and evaluation method of Fig.~3.
  The green line corresponds to the numerically-differentiated compressibility $\kappa$ times the temperature $T_\text{EoS}$ obtained from the EoS-fit of the averaged data.
  The vertical line indicates $\mu (nd^2 =1)$.
  The dashed line corresponds to the on-site density fluctuations $\delta n_0^2 = \langle \hat n^2\rangle - \langle \hat n\rangle^2$ calculated with NLCE for $T_\text{EoS}$. 
  (b)~Comparison of the temperatures $T_\text{FDT}$ (dark red diamonds) and $T_\text{EoS}$ (light red triangles).
  Error bars are the s.e.m.
  }
\end{figure}

In order to extract the temperature from the FDT, we linearly interpolate the compressibility (Sec.~\ref{sec:compressibility}) to match the binning grid of the density fluctuations (Sec.~\ref{sec:binning}) and calculate the local temperature as the ratio between the local variance and compressibility.
The global temperature $T_\text{FDT}$ is calculated as weighted average of the local temperature as a function of the chemical potential for $nd^2 > 0.05$.
In Fig.~\ref{fig:local_temp}, we show the local temperature and compare it to $T_\text{FDT}$ and $T_\text{EoS}$ for $\N = 6$.

\subsubsection{Fluctuations for $\N=3$ and 4}

In Fig.~\ref{fig:fluctuations4} and~\ref{fig:fluctuations3} we show the density fluctuations and the comparison between $T_\text{EoS}$ and $T_\text{FDT}$ for $\N = 4$ and 3 respectively.
In the case of $\N = 4$, we show a dataset with a larger number of frames compared to what has been used in Fig.~3 (35 frames instead of 15) with similar s.e.m. in the total atom number to narrow down the error bar on $T_\text{FDT}$ and allow for a more precise comparison.

\section{Numerical methods}
In this section we present details of the DQMC and NLCE calculations used for results in the main text, including estimates of systematic errors, and a derivation of the HTSE.

\subsection{Determinantal Quantum Monte Carlo}

Averages of the thermal equilibrium observables are evaluated with Determinantal Quantum Monte Carlo (DQMC)~\cite{blankenbeclerMonteCarloCalculations1981,sorellaNovelTechniqueSimulation1989} on $6 \times 6$ lattices by introducing $N(N-1)/2$ auxiliary Hubbard-Stratonovich fields, one for each interaction term~\cite{ibarra-garcia-padillaUniversalThermodynamicsMathrmSU2021a,taieObservationAntiferromagneticCorrelations2022}~\footnote{Previous work applied DQMC for the SU($2N$) Fermi-Hubbard model at half-filling, i.e. $\langle n \rangle=N/2$, using an alternative Hubbard-Stratonovitch decomposition. 
This alternative decomposition is free of the sign problem at half-filling for SU($2N$)~\cite{assaadSpinInvariantAuxiliary1998,wangCompetingOrders2D2014,zhouQuantumMonteCarlo2014,assaadPhaseDiagramHalffilled2005}.}. 
In this method, the inverse temperature $\beta$ is discretized in steps of $\Delta \tau$ (Trotter steps). Results in the main text use $\Delta \tau = 0.05/t$. We obtain DQMC data for 5 different random initial seeds for $U/t = 7.43, 10.38$ and for 20 different random seeds for $U/t =2.34$. 
For each Monte Carlo trajectory we perform $2000$ warm up sweeps through the whole space-time lattice and $8000$ sweeps for measurements.
Furthermore, we use 4 global moves per sweep to ensure ergodicity~\cite{scalettarErgodicityLargeCouplings1991}. 

In addition to statistical errors, (controllable) systematic errors can arise from the finite Trotter step, finite-size effects, and imperfect equilibration of the Monte Carlo algorithm before measurement.
The Trotter error obtained by comparing $\Delta \tau =0.04/t$ and $\Delta \tau =0.05/t$ results is $\lesssim 0.03\, t$ for the temperature and $\lesssim 0.06\,\kB$ for the entropy across all $N$ and $U/t$ considered. 
These correspond to relative errors that are $\lesssim 8\%$, and in all cases are smaller than the statistical error bars~\footnote{For results at $U/t = 7.43$ and $10.38$, relative errors are $\sim 1\%$.}. 
The finite-size error assessed by taking the difference of temperature and entropy obtained from fits using results from $4 \times 4$ and $6 \times 6$ lattices is $\lesssim 0.04\, t$ for the temperature and $\lesssim 0.03\,\kB$ for the entropy for all $N$ and $U/t$ considered, except for $N=6$ at $U/t=2.34$ where the errors are $0.07\,t$ and $0.13\,\kB$. 
The equilibration error is negligible in the homogeneous case and was estimated by comparing results using 2000 and 8000 warm up sweeps. 
This error is $\lesssim 0.002$ for the density and $\lesssim 0.04\, t$ for the energy across all chemical potentials considered in the homogeneous case.

DQMC results are calculated on a grid of $\mu$ and $T$, which can also introduce systematic errors into the fits of the experimental equation of state measurements to theory. 
We use a grid with $\text{d}\mu = 0.25\, t$ and $\text{d}T/t \sim 0.05$ for $T/t \leq 0.6$, and $\text{d}T/t \sim 0.1$ for $0.6 \leq T/t \leq 1.5$. 
Effects due to the coarseness of the $\mu$ grid were estimated by considering differences of temperature and entropy fits using $\text{d}\mu = 0.25\, t$ and $\text{d}\mu = 0.5\, t$ and are $\lesssim 0.007\,t$ and $\lesssim 0.05\,\kB$.

Convergence criteria correspond to the lowest $T/t$ for which fluctuations in entropy for two consecutive $\mu$ data points still fall within statistical error bars.

\subsection{Numerical Linked Cluster Expansion}

Thermodynamic observables are computed using site expansion Numerical Linked Cluster Expansion (NLCE) up to  seventh order (seven sites) for SU($3$), fifth order for SU($4$) and fourth order for SU($6$) Fermi-Hubbard models. 
A brief discussion of the algorithm is presented below. 
Extensive properties of a lattice $\mathcal{L}$ are computed by performing a weighted sum of the property values over all embeddable clusters $c$~\cite{rigolNumericalLinkedCluster2006,tangShortIntroductionNumerical2013}. Formally, 
\begin{equation}\label{eq:nlce_obs}
    P(\mathcal{L})/N_s = \sum_{c\subset \mathcal{L}} L(c) W_P(c)
\end{equation}
where $P(\mathcal{L})$ is the lattice property, $N_s$ is the number of lattice sites, $L(c)$ is the number of ways per site the cluster $c$ can be embedded in the lattice $\mathcal{L}$, and the weights $W_{P}(c)$ are defined as
\begin{equation} \label{eq:nlce_weights}
    W_P(c) = P(c) - \sum_{s\subset c}W_P(s)
\end{equation}
where $P(c)$ is computed by performing exact diagonalization of the Hamiltonian in~\EqMainTextHubbard defined over the cluster $c$. 
The Hilbert space dimension grows exponentially with both system size $N_s$ and number of spin flavors $N$. 
We use spin flavor conservation symmetry to reduce the Hilbert space dimension as mentioned in Ref.~\cite{ibarra-garcia-padillaUniversalThermodynamicsMathrmSU2021a}.
  
Note that Eq.~\ref{eq:nlce_obs} follows from Eq.~\ref{eq:nlce_weights} applied to $c=\mathcal L$. The NLCE works by truncating Eq.~\ref{eq:nlce_obs} to a finite order (finite number of sites in the clusters included), which  yields accurate results when correlation lengths are sufficiently short, as happens when temperature is not too low. 
The NLCE is typically much more accurate than exact diagonalization employing the same number of sites. See the Appendix of Ref.~\cite{ibarra-garcia-padillaUniversalThermodynamicsMathrmSU2021a} for a comparison of bare ED and NLCE for the SU($N$) Hubbard model. 
The NLCE converges to lower temperatures when the density is an integer, and when $U/t$ is larger. 
For both cases, the system has shorter-ranged correlations at a fixed temperature and thus the system properties are captured by small clusters.

The only error in the NLCE arises from truncating Eq.~\ref{eq:nlce_obs} to finite cluster size. We evaluate this by comparing properties calculated with successive orders of the NLCE approximation.  
When two orders of NLCE consistently agree, generally the result is converged and the value equals that in the thermodynamic limit~\cite{tangShortIntroductionNumerical2013}.

The NLCE was computed over $T/t$ grid of $500$ points linearly spaced in the range $[0.5,10]$, i.e. $\text{d}T/t \sim 0.02$.
The $\mu/t$ grid was varied for different $U/t$ and $N$ such that the density, $\expect{n} \in [0,N/2]$. 
For each $U/t$ and $N$, there are $500$ $\mu/t$-points. In the main text (\FigMainTextEoS), the chemical potential is re-scaled with $U$, thus the $\mu$-spacing $\text{d}\mu/U$ varies between $0.01$ and $0.02$. 
The choice of the temperature and chemical potential grids was made to get a dense enough grid to get small fit errors to the experimental data (\FigMainTextEoS).

\subsection{High Temperature Series Expansion}

We also present results from a second-order high temperature series expansion (HTSE) in $t/T$ which is accurate for $T \gtrsim t$. 
In the following, subscripts are used to indicate the expansion's order in $t/T$ for the different physical quantities $\expect{O}_\ell$.

\subsubsection{Zeroth-order HTSE ($t=0$)}
The energy as a function of the occupation of a single site in the grand canonical ensemble is given by,
\begin{equation}
    \epsilon_0(n) = \frac{U}{2}n(n-1) - \mu n + g \delta_{n,\alpha},
\end{equation}
where we introduced the last term to allow extraction of the $p_\alpha$ by differentiating the partition function with respect to $g$.
The partition function for the single site is given by $z_0 = \sum_{n=0}^N \binom{N}{n} e^{-\beta \epsilon_0(n)}$, and the grand canonical free energy per site is $\Omega_0 = -T \ln z_0$. For compactness, we define $y = e^{-\beta U}$, $x = e^{\beta \mu}$, and $w = e^{-\beta g}$.

In the zeroth-order HTSE $\expect{p_\alpha}_0 = \lim_{g \to 0} \frac{\partial \Omega_0}{\partial g}$, which gives
\begin{equation}
    \expect{p_\alpha}_0 = \frac{ \binom{N}{\alpha} y^{\frac{1}{2} \alpha (\alpha -1)} x^\alpha}{\sum_{n=0}^N \binom{N}{n} y^{\frac{1}{2} n (n-1)} x^n}. 
\end{equation}
The density $\expect{n}_0$ and number of on-site pairs $\expect{\mathcal{D}}_0 = \expect{n(n-1)/2}_0$ are related to $\expect{p_\alpha}_0$ via $\expect{n}_0 = \sum_{\alpha = 1}^N \alpha \expect{p_\alpha}_0$, and $\expect{\mathcal{D}}_0 = \sum_{\alpha = 2}^N \binom{\alpha}{2} \expect{p_\alpha}_0$.

\subsubsection{Second-order HTSE}

The first correction to the free energy is second-order in $t/T$, i.e. $\Omega_2 = \Omega_0 + \Delta \Omega$~\cite{oitmaaSeriesExpansionMethods2006}, where $\Delta \Omega$ is presented in Refs.~\cite{hazzardHightemperaturePropertiesFermionic2012,taieSUMottInsulator2012b,hofrichterDirectProbingMott2016} for $g=0$. For general $g$ we find
\begin{widetext}
\begin{align}\label{eq:CF_g0}
    -\beta \Delta \Omega = q N\left(\frac{\beta t}{z_0} \right)^2 \Bigg[&\frac{1}{2} \sum_{n=1}^N \binom{N-1}{n-1}^2 x^{2n-1} y^{(n-1)^2}w^{\delta_{\alpha,n} +  \delta_{\alpha,n-1}}  \nonumber \\
    &- \frac{1}{\beta U} \sum_{n \neq m} \binom{N-1}{n-1}\binom{N-1}{m-1} \frac{x^{n+m -1} y^{\frac{1}{2} \left[n(n-1) + (m-1)(m-2)\right]} w^{\delta_{\alpha,n} +  \delta_{\alpha,m-1}}}{n-m}\Bigg],
\end{align}
\end{widetext}
where $q$ is the coordination number.
Following the same procedure as before, $\expect{p_\alpha}_2 = \expect{p_\alpha}_0 + \expect{\Delta p_\alpha}$, where $\expect{\Delta p_\alpha}$ is given by $\expect{\Delta p_\alpha} = \lim_{g \to 0} \frac{\partial \Delta \Omega}{\partial g}$, yielding
\begin{widetext}
\begin{align}
    \expect{\Delta p_\alpha} = q N\left(\frac{\beta t}{z_0} \right)^2 \Biggl[& \frac{1}{2} \sum_{n=1}^N \binom{N-1}{n-1}^2 x^{2n-1} y^{(n-1)^2} \big(-2 \expect{p_\alpha}_0 + \delta_{\alpha,n} + \delta_{\alpha,n-1} \big) \nonumber \\
    & - \frac{1}{\beta U} \sum_{n \neq m} \binom{N-1}{n-1}\binom{N-1}{m-1} \frac{x^{n+m -1} y^{\frac{1}{2} \left[n(n-1) + (m-1)(m-2)\right]}}{n-m}  \big(-2 \expect{p_\alpha}_0 + \delta_{\alpha,n} + \delta_{\alpha,m-1} \big) \Biggl].
\end{align}
\end{widetext}
The second order expectations $\expect{n}_2$ and $\expect{\mathcal{D}}_2$ are related to $\expect{p_\alpha}_2$ via $\expect{n}_2 = \sum_{\alpha = 1}^N \alpha \expect{p_\alpha}_2$, and $\expect{\mathcal{D}}_2 = \sum_{\alpha = 2}^N \binom{\alpha}{2} \expect{p_\alpha}_2$, respectively.

\bibliography{references}